\documentclass{aa}  
\usepackage{rotating}
\usepackage{amsmath}
\usepackage{float}
\usepackage{multirow}
\usepackage{multicol}
\usepackage{adjustbox}
\usepackage{graphicx}
\usepackage{txfonts}
\begin{document}

   \title{Tracing the outflow kinematics in Type 2 Active Galactic Nuclei}

   \author{ Jelena Kova\v cevi\' c-Doj\v cinovi\'c
   \inst{1},
         Ivan Doj\v cinovi\'c
         \inst{2},
         Ma\v sa Laki\' cevi\' c
         \inst{1},
          \and
          Luka \v C. Popovi\' c\inst{1,3}
          }

   \institute{Astronomical  Observatory,  Volgina  7, 11060  Belgrade, Serbia\\
              \email{jkovacevic@aob.bg.ac.rs}
         \and           
       Faculty of Physics, University of Belgrade, Studentski Trg 12, Belgrade, Serbia\\
                  \and
     Faculty of Mathematics, University of Belgrade, Studentski Trg 16, Belgrade, Serbia\\
          }



  \abstract
{

We have used the sample of 577 active galactic nuclei Type 1.8-2 spectra (z < 0.25), taken from Sloan Digital Sky Survey, to trace the  influence of the outflow kinematics to the profiles of different emission lines (H$\beta$, [O III], H$\alpha$, [N II], [S II]). All considered lines were fitted with two Gaussian components: one which fits the core of the line, and another which fits the wings. We gave the procedure for decomposition of H$\alpha$+[N II] wavelength band, for the spectra where these lines overlap. The influence of the gravitational/non-gravitational kinematics to the line components is investigated by comparing the dispersions of the line components with stellar velocity dispersion. We found that wing components of all considered emission lines have pure non-gravitational kinematics, the core components are  consistent with gravitational kinematics for the H$\alpha$, [N II] and [S II] lines, while in the [O III] there is evidence for contribution from non-gravitational kinematics. We adopted the wing components as proxy of the outflow contribution and we investigated the outflow kinematics by analysing the correlations between widths and between shifts of the wing components of different lines. For this purpose, we used the subsets in which wing components are detected in both compared lines, and could be fitted independently. We found the strong correlations between shifts and between wing component widths of all considered lines, with exception of the H$\beta$ wing component width. These correlations indicate that outflow dynamics systemically affects all emission lines in spectrum. However,  it reflects with different strength in their profiles, which is observed as different widths of the wing components. This is investigated by comparison of the mean widths of the wing components in subsets where wing components are present in all lines. The  strongest outflow signature is observed in the [O III] lines, which have the broadest wing components, weaker in H$\alpha$ and [N II], and the weakest in [S II]. These results imply that considered lines arise in different parts of an outflowing region.

}

    \titlerunning{Tracing the outflow kinematics in Type 2 AGNs}
               \authorrunning{Kova\v cevi\' c-Doj\v cinovi\'c et al.}                                      

   \maketitle
%

\section{Introduction}\label{1}

Active galactic nuclei (AGNs) are one of the most powerful energy
sources in the Universe. The enormous amount of energy radiated from AGN central source is formed in process of accretion around super-massive black hole, that is followed by the fast motion of the emitting gas and appearance of outflows.
The large-scale phenomena, gas outflows, which origin is still unclear, have been observed in all wave bands of AGN spectra, from the radio, IR, optical, UV, to the X band \citep[for a review see][]{Veilleux2005}. It is proposed that outflows can be a channel for AGN feedback, which could be the explanation for the observed correlation between central black hole mass and host galaxy properties \citep{DiMatteo2005, Fabian2012, Harrison2018}.
Several models have been proposed as driving mechanisms of the AGN outflows: radiation pressure \citep{Osterbrock2006, Ishibashi2015}, interaction of the radio jet with clouds \citep{Nesvadba2008, Mukherjee2018}, small-scale AGN winds \citep{Tombesi2015, Costa2020}, hybrid models \citep{Everett2005}, etc. The  radiation pressure on dust, as driving mechanism for galactic-scale AGN outflows, is investigated from some theoretical \citep{Ishibashi2015} and observational point of view \citep{Zakamska2016}. Several studies found the connection between the level of radio emission in AGN host galaxies and the presence of ionized outflows \citep{Mullaney2013, Zakamska2014, Molyneux2019, Jarvis2021}. Also, observational evidences that outflow kinematics and radio jets seem to be related are reported \citep{Smirnova2007, Nesvadba2017, Berton2021}. However, \cite{Wang2018} found no connection between the radio properties and the outflow kinematics.
 The main driver of AGN outflows remains major topic of debate in the literature and further investigation of the outflow origin and kinematics should give us an insight in complex AGN physics and evolution.

 In this study, we focused on ionised outflows detected with optical emission-lines. The spatially resolved spectroscopic observations significantly contributed to better understanding of the outflow kinematics and physics. The size of outflow is found to be up to several kpc \citep{Harrison2014, Karouzos2016a, Kang2018},  which is comparable to the size of the galaxy bulge. Except the size, the outflow velocity, extension of the ejected material and its mass, as well as  associated energy, are estimated for the particular objects \citep{Harrison2014, Karouzos2016b, Kang2018}. The outflow velocity is estimated to be generally up to $\sim$ 1000 km s$^{-1}$, and in extreme cases up to $\sim$  2000 km s$^{-1}$ \citep[see][]{Bae2016, Komossa2018}. It is proposed that outflow geometry can be represented with the biconical outflow models in combination with extinction material \citep{Crenshaw2010, Bae2016}.

Several studies concluded that the profiles of narrow lines are superposition  of two components: one arising in gas which follows the gravitational kinematics of the stellar motion, and the other which represents the non-gravitational kinematics, originating in gas outflow. This is confirmed by numerous spatially resolved observations of particular objects \citep{Smirnova2007, Karouzos2016a, Bae2017, Kang2018}, as well as by the systematic studies using the large samples of spatially integrated spectra \citep[see e.g.][]{Nelson1996, Mullaney2013, woo2016, Rakshit2018}. All mentioned studies analysed the low redshift AGNs (z$<$0.4). It is found that the gravitational component  dominates in the core of the emission line, while outflow kinematics is dominantly traced in wing component, usually represented as one Gaussian, broader than the core component, often shifted to the blue or to the red relative to the core component. 
\cite{woo2016} found that wing component is present in 44\% of the AGN Type 2 spectra.
However, it seems that  the outflow contribution could be partly present in core component as well \citep{Karouzos2016a, woo2016, Sexton2021}. Moreover, several studies point out that total emission line profile, including the core component, could be affected by outflow emission \citep{Liu2013, Zakamska2014, Jarvis2019, Davies2020a}. \cite{Venturi2021} found that the broadness of the line profiles might be also associated with turbulent gas.

The [O III] emission lines are found to be very good tracer of the AGN driven outflows, and they are most studied lines related to outflow kinematics. The results of the comparable analysis of the outflow influence to the [O III]  and H$\alpha$ profiles indicate  stronger outflow influence to the [O III] lines \citep{Karouzos2016a, Bae2017, Kang2017, Kang2018}. In both lines, the outflow kinematics is dominantly present in wing component, while the core component represent gravitational kinematics in  H$\alpha$, and mixture of the  gravitational and outflow kinematics in the case of the [O III] lines \citep{Karouzos2016a, woo2016, Sexton2021}. \cite{Kang2017} analysed the total line profiles of these lines in a large sample of AGNs Type 2 and found that outflow kinematics, reflected in the width and velocity shift of the line profiles, correlates in H$\alpha$ and [O III] lines. However, they conservatively excluded from the analysis all objects which are candidates to be hidden AGNs Type 1, i.e. could possible have broad H$\alpha$ component in blended [N II]$+$H$\alpha$ wavelength band \citep[see][]{woo2014, Oh2015, Eun2017}.

Generally, the manifestation of the outflow kinematics is less investigated on the other lines beside the [O III]. One of the reasons for that can be in difficulty of decomposition of the blended [N II]$+$H$\alpha$ wavelength band in AGNs Type 2, where the strong wing components of these three lines can overlap and form the fake, pseudo-broad component. In fact, without very careful spectroscopic analysis, one cannot be sure is the blended [N II]$+$H$\alpha$ in AGNs typically classified as Type 2, the sum of the three strong wing components, true hidden broad H$\alpha$ component \citep[see][]{woo2014, Oh2015, Eun2017} or mixture of all these together. The problem of the correct decomposition of the complex [N II]$+$H$\alpha$ certainly complicates the research of the outflow kinematics in the H$\alpha$ and [N II] lines, specially their independent analysis, since these lines are often fitted with some fitting constrains \citep{Karouzos2016a, Bae2017, Luo2019, Davies2020b}.
On the other hand, the [S II] and H$\beta$  lines are usually weak, and their wing components are in level of noise in majority of the spectra.

 The aim of this work is to use the multiple emission lines ([O III], H$\beta$, H$\alpha$, [N II] and [S II]) to study the outflow kinematics in the low redshift sample of AGNs Type 1.8-2. We investigated if there is a difference between considered lines related to gravity/outflow influence to their line profiles, and we searched for the correlations between their kinematical properties in order to trace the outflow influence. Unlike the study of \cite{Kang2017}, here we have been especially focused on making the procedure for decomposition of the blended [N II]$+$H$\alpha$, and distinction of the sum of the wing components from possibly present hidden broad H$\alpha$, in order to analyse the outflow kinematics in these spectra. Since in this research we investigate the outflow influence to some of the weak emission lines,  we chose the sample of several hundreds of Type 1.8-2 AGN Sloan Digital Sky Survey (SDSS) spectra but with highest possible signal-to-noise ratio (S/N). 
 In this way, we could perform very precise and sophisticated fitting procedure of the complex [N II]$+$H$\alpha$ wavelength band and trace the weak outflow contribution in the H$\beta$ and [S II] lines.

In Section \ref{2} we describe the sample selection and fitting procedure. The results are shown in Section \ref{4}. In Section \ref{5} we discuss the obtained results and in Section \ref{6} we outline the conclusions of this research.  Throughout this
paper we used following cosmological parameters: $H_0=70$ km s$^{-1}$ Mpc$^{-1}$, $\omega_m=0.30$ and $\omega_{\Lambda}=0.70$.

\section{Data and method of analysis}\label{2}

\subsection{The sample selection}\label{2.1}

In order to choose a sample of AGN Type 2 spectra with highest possible S/N, we searched the SDSS database\footnote{\url{https://www.sdss.org/dr14/}} using the  Structural Query Language (SQL) search of the Data Release 14 \citep[DR14, see][]{Abolfathi2018, Blanton2017}. It is the second data release of the fourth phase of the Sloan Digital Sky Survey (SDSS-IV). It contains the optical single-fiber spectroscopy observations through July 2016, done with 2.5 m Sloan Foundation Telescope at the Apache Point Observatory. The details about SDSS spectrographs and spectral resolution can be found in \cite{SM2013}.

Using the SQL search in SDSS DR14 database we chose the galaxy spectra which satisfy the following criteria:

\begin{enumerate}

\item  objects to be spectral class 'galaxy',

\item  to belong to AGN part of BPT diagram \citep[see][]{Baldwin81},

\item median S/N per pixel of the whole spectrum $>$ 20,

\item equivalent widths (EWs) of the emission lines: H$\beta$, [O III]$\lambda\lambda$ 4959, 5007 \AA, H$\alpha$, [N II]$\lambda\lambda$ 6548, 6583 \AA \  to be larger than 5 \AA. 
\end{enumerate}

\begin{figure} 
 \centering
\includegraphics[width=80mm]{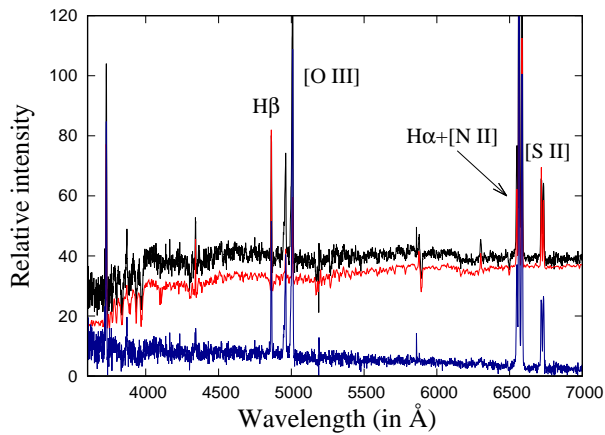}
\caption{ Decomposition of the spectrum (SDSS J105621.57-005320.4) to the host galaxy and AGN contribution using spectral principal component analysis (SPCA). Black line - observed data,  red line - host galaxy contribution obtained from SPCA, blue line  - AGN contribution (host galaxy contribution subtracted from observed spectrum).
\label{fig1}}
 \end{figure}

\begin{figure} 
 \centering
\includegraphics[width=70mm]{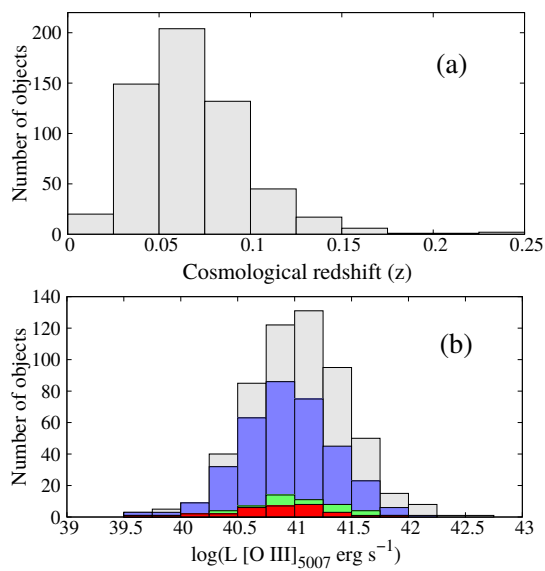}
   \caption{ (a) Histogram of distribution of cosmological redshift for total sample. (b) Histogram of distribution of [O III]$\lambda$ 5007 \AA \ luminosity for total sample (grey bars) and for different subsets ('unblended' subsample - blue bars, objects from 'unblended' subsample with detected wing components in both, H$\alpha$ and [N II] lines - green bars, and the same just for [N II] and [S II] lines - red bars.)
    \label{fig0}}
 \end{figure}

In order to get more confident AGN classification, we cross-match two independent BPT classifications, one given in \cite{Bolton2012} ('AGN' subclass) and the other one given in \cite{Thomas2013} (available et Emission Line Port SDSS Table)\footnote{ The procedures for computations of galaxy properties given in \cite{Bolton2012} and \cite{Thomas2013} (primarily made for DR9) have been applied to all DR14 objects that the spectroscopic pipeline classifies as a galaxy. The measurements are available in SDSS Tables.}. They are using different codes for measurement of emission line fluxes needed for BPT diagram and different separation curves in order to detect AGNs among galaxies. 
We found matching in 80\% of objects classified as Seyfert 2 AGNs using these two approaches, and we adopted only the objects which are classified as Seyfert 2 in both cases. 

Using above criteria we obtained the spectra of 588 AGNs, predominantly Type 2, but also with possible presence of the AGNs Type 1.9 and Type 1.8. Namely, in some cases, objects can be identified as AGNs Type 1.8/1.9 only after subtraction of the strong host galaxy contribution, which dominates over weak AGN continuum and flux of broad H$\alpha$, or after applying careful spectral decomposition which reveal broad H$\alpha$ in blended H$\alpha+$[N II].

Three spectra were excluded from the sample due the bad pixels near H$\alpha$ and [S II]. Eight objects have double peak narrow lines which make them candidates for the binary black hole system. Since this system is very complex and out of scope of this work these objects were excluded from the final sample. This left 577 objects in the sample.

The spectra were corrected for the Galactic reddening using the standard extinction law given in  \cite{ho1983} and  extinction coefficients given in \cite{SF2011}, available from the NASA/IPAC Infrared Science Archive 
(IRSA)\footnote{\url{http://irsa.ipac.caltech.edu/applications/DUST/}}. Afterwards, the spectra were corrected for the cosmological redshift measured with  automated procedure described in \cite{Bolton2012}. In that procedure, the redshift of each galaxy is determined using best fit with redshifted template basis made of linear combination of 4 leading galaxy eigenspectra. 

We applied the Spectral Principal Component analysis (SPCA) in order to decompose the spectrum to the  host galaxy component, and to the pure AGN contribution. 
This method is performed as in  \cite{2006AJ...131..84}, using the 10 QSO eigenspectra and 5 galaxy eigenspectra  \citep[see][]{yip04a, yip04b}.  We perform non-parametric fitting of the observed spectra with linear combination of 15 eigenspectra. The host galaxy contribution is obtained as the linear combination of the 5 galaxy eigenspectra obtained from the best fit. We masked all narrow emission lines in obtained host-galaxy spectra, and subtracted it from  the observed spectra. In this way, we obtained the pure spectra of AGNs. The example of the host galaxy/AGN decomposition is shown in Fig.~\ref{fig1}. The weak continuum emission is removed in all spectra, using the continuum windows from  \cite{ku2002}.

 \subsection{The sample properties and selection effect}\label{2.2}

The histograms of distribution of cosmological redshift and [O III]$\lambda$ 5007 \AA \ luminosity (L) for the sample of 577 AGNs are shown in Fig. \ref{fig0}  (grey bars). Using described selection criteria we obtained low-redshift AGN sample (z$<$0.25) in which 95\% of objects have z$<$0.125 and log(L[O III]$_{5007}$ erg/s) within the range [40.25--42]. 

We tested the influence of the selection criteria to the properties of the final sample. We found that the criterium of the line strength (criterium number 4, EWs of considered lines to be larger than 5 \AA) has no significant influence to the redshift distribution of the sample, while deleting of this criterium leads to the larger number of objects with smaller luminosities of L[O III]$_{5007}$. On the other hand, the decrease of the lower limit for S/N (criterium number 3), leads to the larger number of objects with slightly higher redshift and [O III] luminosity.

We conclude that criteria of the lower limit for S/N and lower limit for the line strength lead to losing the spectra with slightly higher redshifts (z$>$0.125) and with broader luminosity range. Although we lose possibility to investigate the outflow in larger diversity of objects, the criteria of high S/N and line strength are required for this investigation, since they enable the analysis of the asymmetries of the weak emission lines as H$\beta$ and [S II], which are diluted by the noise in spectra with lower S/N. Therefore, the chosen sample consists of the low-redshift AGNs with high-quality spectra and prominent emission lines, in which we can analyse the emission line shapes very precisely.

\subsection{The fitting procedure}\label{2.3}

 We used double-Gaussian or single-Gaussian model for emission line decomposition, and we applied a $\chi^2$ minimalization routine to obtain the best fit parameters, similar as in the case of Type 1 AGNs in \cite{popovic2004} and \cite{kovacevic2010}. Double-Gaussian model implies one Gaussian which fits the core of the line (core component), and one which fits the wings (wing component), which is often used to describe the narrow emission line shapes \citep[see e.g.][etc.]{Mullaney2013,woo2016,Sexton2021}.

\subsubsection{Decomposition of the [O III], H$\beta$ and [S II] lines}\label{2.3.1}

Since the [O III], H$\beta$ and [S II] lines do not overlap with other lines in spectra,  they were fitted completely independently through whole sample, using double-Gaussian (wing$+$core) model, where wing and core Gaussian components of each line have the free parameters for width, shift and intensity. 
The only parameter constrains for these lines is done within doublets. The [O III] doublet is fitted to both doublet lines ([O III]$\lambda$5007 \AA \ and [O III]$\lambda$4959 \AA) have the same width and shift of the core (and wing) Gaussian, and line ratio to be 2.99 \citep[see][]{dim07}. The core and wing components of the [S II]$\lambda\lambda$ 6717, 6731 \AA \  doublet are fitted with same widths and shifts for both doublet lines, but their intensity ratio is not fixed. In 16 objects, after removing the host galaxy contribution, we observed the weak broad H$\beta$, originating from the Broad Line Region (BLR), which is fitted with one broad Gaussian. 

\begin{figure} 
 \centering
\includegraphics[width=80 mm]{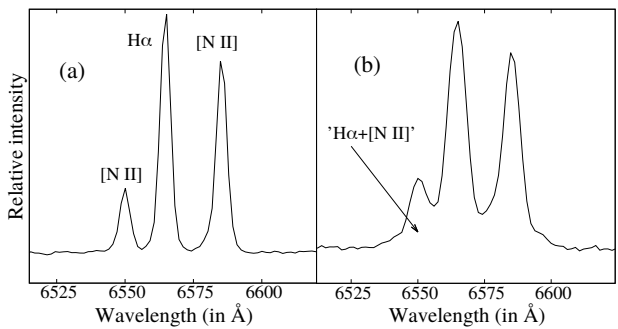}
 
 \caption{ The [N II]$+$H$\alpha$ wavelength band in spectra which belong to (a) 'unblended' and (b) 'blended' subsample (SDSS J100551.19+125740.6 and SDSS J112408.63-010927.9, respectively). 
\label{fig1_1}}
 \end{figure}

We found that in some spectra wing components of [O III], [S II] and H$\beta$ have  widths comparable to the widths of the core components, and their contribution to the line shape is very poor. Therefore, we adopted two criteria in order to avoid fake wing components: (i) The wing component amplitude-to-noise ratio (A$_w$/N) to be larger than 3. The wing components with A$_w$/N $<$ 3  were neglected. (ii) For the wing components which satisfy the first criterium,  we applied the extra-sum-of-squares F-test \citep{Lupton1993}, in order to check the justification of their implementation in model, similarly as in \cite{Sexton2021}. This test compares the improvement of sum-of-squares with the more complicated model vs. the loss of degrees of freedom. The test is done for wing components of each line separately, and wing components are adopted if P-value $<$ 0.05, for null hypothesis that there is no significant improving of the fit by applying the model which includes more parameters, comparing the fit with model with smaller number of parameters. The wing components for which  P-value is larger than 0.05 were neglected since one cannot be sure in their presence. In cases where wing components were  neglected, the lines were fitted with single-Gaussian model.

 We found that adopted double Gaussian and single Gaussian models describe well line shapes of majority of the [O III], H$\beta$ and [S II] lines. However, in $\sim$2\% of objects from the sample, we noticed the complex shapes of the [O III] lines, and in some cases H$\beta$ lines as well. We excluded from correlation analysis 12 the most extreme cases, for which we found by visual inspection that their complex line shapes could not be fitted with adopted models. These objects were analysed separately in Appendix \ref{A}.

\subsubsection{The division to the 'unblended' and 'blended' subsamples}\label{3.2}

To correctly decompose [N II]$+$H$\alpha$ wavelength band, we adopted following procedure: we divided whole sample into two subsamples -- one which consists of spectra where [N II] and H$\alpha$ lines are unblended, i.e. do not overlap each other (hereafter 'unblended' subsample), and to the other one where [N II]$+$H$\alpha$ wavelength band is blended (hereafter 'blended' subsample). The blended [N II]$+$H$\alpha$  wavelength band in fact can be (a) sum of the broaden wings of the [N II]$+$H$\alpha$ lines which form pseudo-broad component, (b) true H$\alpha$ BLR component, or in the most complex cases, (c) all these together. 

To form the 'unblended' subsample, we single out all spectra from total sample in which emission line flux at $\sim$ 6575 \AA \ (minimum between H$\alpha$ and [N II] 6583 \AA \ line) is smaller than 5\% of [N II] 6583 \AA \ line intensity. In this way we obtained the 346 spectra, in which H$\alpha$ and [N II] 6583 \AA \ lines do not overlap or overlap very slightly, i.e. flux at $\sim$ 6575 \AA \ (Flux$_{6575}$) is close to the continuum level. We found that in 80\% of objects from this subsample, the Flux$_{6575}$ is in the level of the noise (Flux$_{6575}$/N $<$ 3). The rest of the spectra (219 objects, after excluding 12 objects with the complex [O III] shape) belong to the 'blended' subsample. Following this separation criterium, the 'unblended' spectra make $\sim$60\%, and 'blended' $\sim$40\% of the total sample. The examples of the [N II]$+$H$\alpha$ wavelength band in spectra from both subsamples are given in Fig. \ref{fig1_1}. 

We first performed the decomposition of the H$\alpha$ and [N II] lines in 'unblended' subsample, and afterwards, we used the empirical relationships found in 'unblended' subsample for establishing the decomposition procedure for 'blended' subsample (see Sec. \ref{4.3}). 
The H$\alpha$ and [N II] lines in 'unblended' subsample are fitted independently, with double Gaussian model, without any parameter constrains for the core and wing components. Therefore, they are fitted in the same way as the [O III], H$\beta$ and [S II] lines in the total sample. The decomposition of the [N II]$\lambda\lambda$ 6548, 6583 \AA \  doublet lines is done similarly as for [O III] doublet: both doublet lines ([N II]$\lambda$6548 \AA \ and [N II]$\lambda$6583 \AA) have the same width and shift of the core (and wing) Gaussian, and ratio of component intensities 3.05 \citep[see][]{Dojcinovic2022, Osterbrock2006}.  Similarly as for the [O III], [S II] and H$\beta$ lines, the wing components of the  H$\alpha$ and [N II] lines in 'unblended' subsample were adopted only if they satisfy criteria of  A$_w$/N $>$ 3 and F-test (P-value $<$ 0.05). In the case if wing components do not satisfy some of these criteria, the H$\alpha$ and [N II] were fitted with single Gaussian model.

\begin{figure*}
 \centering
\includegraphics[width= 170mm]{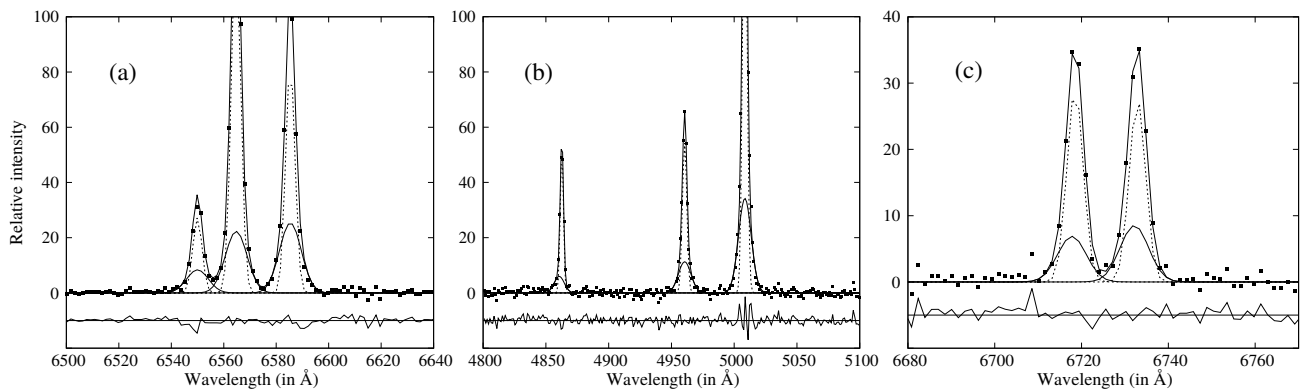}
  
\caption{ Example of fit decomposition for object SDSS J125025.82+001954.9 from 'unblended' subsample: (a) H$\alpha$ and [N II], (b) H$\beta$ and [O III] and (c) [S II]. The core components are denoted with dotted line and wing components with solid line.
\label{fig2}}
 \end{figure*}

The example of the fit of all considered lines for spectrum from 'unblended' subsample is shown in  Fig.~\ref{fig2}, while the decomposition of the 'blended' subsample is presented in Sec. \ref{4.3}.

\subsection{Measuring the spectral parameters}\label{3.3}

 In this investigation we analyse the gas kinematics using the velocity shift and width (dispersion) of the separated core and wing Gaussian components, obtained from the best fit.  The fitting procedure is done after systemic correction of spectra for cosmological redshift  \citep[determined using galaxy eigenspectra, see][]{Bolton2012}, and the velocity shifts of components are measured relative to the transition wavelength in rest-frame spectra. 
 
 The uncertainties of the line fitting parameters were estimated for each line in each spectrum by applying Monte Carlo method. We used the measured values from the  emission-line fits to construct the model, and then we made 100 mock spectra for each source by adding the random noise to model spectra. The random noise was limited to do not exceed the measured 3$\sigma$ noise for each object. After we obtained the fitting parameters of the mock spectra,  we took the 1$\sigma$ dispersion of the parameters as the fit uncertainty.
The lines with small intensities, [S II] and H$\beta$, are found to have larger uncertainties comparing the other considered lines in spectra. The mean shift and width uncertainties for wing components of [S II] are 46 km s$^{-1}$ and 37 km s$^{-1}$, for H$\beta$ are 41 km s$^{-1}$ and 30 km s$^{-1}$, while for  [N II], H$\alpha$ and [O III] these values are in the range of 9-13 km s$^{-1}$. The mean shift/width uncertainties for the core components of all considered lines, as well as for single Gaussian lines, are in the range 1-8 km s$^{-1}$.

The widths of the wing Gaussian components are corrected for instrumental SDSS width, following the formula: $\sigma_{cor}=(\sigma_{obs}^2 - \sigma_{inst}^2)^{1/2}$, where $\sigma_{obs}$ is Gaussian width measured from spectrum, $\sigma_{inst}$ is instrumental SDSS width, and $\sigma_{cor}$ is corrected dispersion, which is used in analysis.
SDSS provides the $\sigma_{inst}$ measured for different lines for each object, which we used for data correction. In this sample the instrumental widths are within interval 48 - 83 km s$^{-1}$.

 The stellar velocity dispersions  ($\sigma_{*}$) are obtained from SDSS \citep{Thomas2013}. The measurements are based  on adaptations of the publicly available codes GANDALF v1.5 \citep{Sarzi2006} and pPXF \citep{Cappellari2004}, which fit stellar population and evaluate the stellar kinematics. Although this procedure is primary applied for the BOSS massive galaxies \citep[z > 0.15, see][]{Dawson2013}, \cite{Thomas2013} demonstrated that $\sigma_*$ estimates using this algorithm are reliable for general galaxy sample. The $\sigma_{*}$ of SDSS spectra were extracted with a 3'' aperture and they may be affected with rotational broadening and inclination effect. All $\sigma_{*}$  below the SDSS instrumental resolution of 70 km s$^{-1}$ were excluded from analysis.

\section{Results}\label{4}

 \subsection{The line profiles: gravitational vs. non-gravitational contribution}\label{4.0}

\begin{figure*} 
 \centering
\includegraphics[width=160mm]{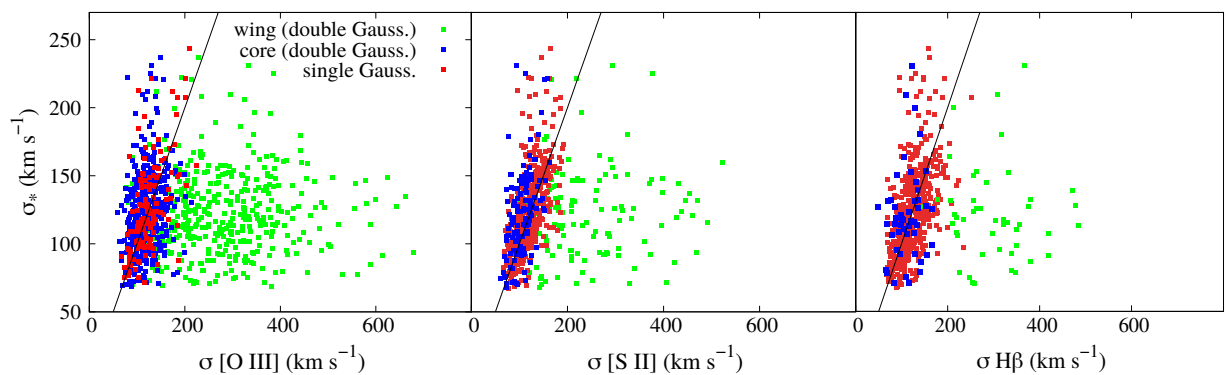}
\caption{Correlations between the $\sigma_{*}$ and widths of single Gaussian lines (red dots), wing and core components of the double-Gaussian lines (green and blue dots), for [O III], [S II] and H$\beta$ lines (total sample). The one-to-one relation is denoted with solid line. \label{fig2_1}}
 \end{figure*}

\begin{figure} 
 \centering
\includegraphics[width=65mm]{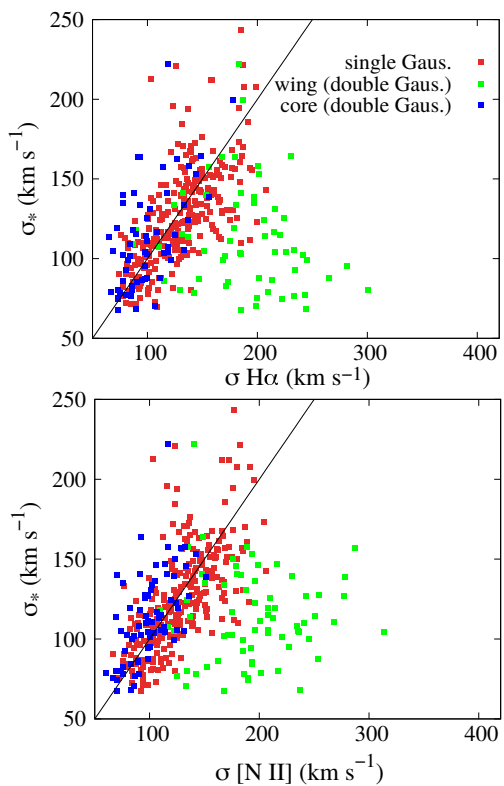}
\caption{The same as in Figure \ref{fig2_1}, just for H$\alpha$ and [N II] lines. The correlations are done for 'unblended' subsample, where H$\alpha$ and [N II] lines are fitted independently. The one-to-one relation is denoted with solid line. \label{fig2_1_1}}
 \end{figure}

Although the  double Gaussian model (core$+$wing component) is frequently applied in narrow emission line decomposition, there is still concern about physical meaning of that decomposition \citep[see e.g.][]{Zhang2011, Davies2020b}. The interesting questions are: do wing and core components dominantly represent emission from different emission regions, and is it justified to analyse them separately? Is it the same for all narrow  emission lines?

The previous investigations which analysed the influence of the gravitation to the narrow emission line profiles mostly compared the total line profile width versus $\sigma_{*}$ \citep[][etc.]{Greene2005a, woo2016, woo2017, Kang2017}, where $\sigma_{*}$ is assumed to be a good indicator of the gravitation motion. The widths of the wing and core components  were rarely used separately in correlation analysis with  $\sigma_{*}$, except  in few studies of AGNs Type 1, only for the [O III] lines \citep{Xiao2011, Sexton2021}. 
 
Here we investigated the influence of the gravitational motion to the widths of the separated wing and core components of different lines. We try to check do these components dominantly originate in kinematically different gas in our sample, and whether there is a difference between considered lines related to gravity/outflow influence. Therefore, we plotted the widths of the core and wing components vs. $\sigma_{*}$ for lines decomposed with double Gaussian model. Additionally, in the same graphs we added the widths of single Gaussian lines vs. $\sigma_{*}$. In Fig. \ref{fig2_1} we showed the plots for the [O III], [S II] and H$\beta$ lines for total sample, and in Fig. \ref{fig2_1_1} for H$\alpha$ and [N II] lines for 'unblended' subsample, where these lines are fitted independently, without fitting constrains.

\begin{table}
\begin{center}
\caption{Correlations between the stellar velocity dispersion ($\sigma_{*}$) and velocity dispersions of the core components of double Gaussian lines and single Gaussian lines. \label{T3}}
\begin{tabular}{|c | c c c|}
\hline

 & &  \multicolumn{2}{c|}{$\sigma_*$}   \\
\hline
\hline
 & &  core (double Gauss.) & single Gauss.   \\ [1ex]

    \multirow{2}{*}{  $\sigma_{[O III]}$ }&r  & 0.26 & 0.58

	\\ 
    &P & 1.7E-7 &  9.1E-13

\\ [1ex]
\hline
  \multirow{2}{*}{  $\sigma_{[N II]}$}&r  & 0.58 & 0.65

	\\
    &P & 1.5E-7  & 0

\\  [1ex]

 \hline

  \multirow{2}{*}{  $\sigma_{H\alpha}$ }&r  & 0.58  & 0.65

	\\
    &P & 1.1E-6   & 0

\\  [1ex]
\hline
 \multirow{2}{*}{  $\sigma_{[S II]}$ }&r  & 0.49 & 0.63

	\\
    &P & 1.2E-9 & 0

\\  [1ex]

\hline
\multirow{2}{*}{  $\sigma_{H\beta}$}&r  & 0.16 & 0.52

	\\
    &P & 0.29 & 0

\\   [1ex]
\hline

\end{tabular}
\tablefoot{For the [O III], [S II] and H$\beta$ lines, correlations are calculated using total sample, while for H$\alpha$ and [N II] they are calculated using 'unblended' subsample.

}
\end{center}
\end{table}

We found that the widths of wing components of the [O III], [S II], H$\alpha$, [N II] and H$\beta$ lines (denoted with green dots in Fig. \ref{fig2_1} and Fig. \ref{fig2_1_1}) show no correlation with $\sigma_{*}$, which implies their non-gravitational origin in all these lines. On the other hand, the widths of the core components of the double-Gaussian lines (blue dots in Fig. \ref{fig2_1} and Fig. \ref{fig2_1_1}) are in significant correlation with $\sigma_{*}$ for the [S II], H$\alpha$ and [N II] lines, which implies their dominantly gravitational origin. However, the widths of the [O III] core components are in weak correlation with $\sigma_{*}$, while no correlation is found between the widths of H$\beta$ core components and $\sigma_{*}$. The coefficients of correlations of different lines versus  $\sigma_{*}$  are given in Table \ref{T3}. The weak correlation of the [O III] core components width with $\sigma_{*}$ implies that beside gravitational influence, the [O III] core components are probably affected with non-gravitational influence in our sample, which is in accordance with  previous results \citep{Karouzos2016a, woo2016, Sexton2021}. Similarly, the lack of correlation between the width of H$\beta$ core components and $\sigma_{*}$  could be caused by strong non-gravitational influence in the core component as well, but also it could be partly  caused by the small number of detected double Gaussian H$\beta$ lines in the sample. Contrary to the core components, the significant correlations are found for the widths of the single Gaussian [O III] and H$\beta$ lines versus $\sigma_{*}$. Generally, we found that the widths of the single Gaussian emission lines (red dots in Fig. \ref{fig2_1} and Fig. \ref{fig2_1_1}) are in stronger correlations with $\sigma_{*}$ comparing the core components of the same kind of lines (which are fitted with double Gaussian model).  We noticed that for $\sigma_{*}$ > 170 km s$^{-1}$, stellar velocity dispersions become slightly larger than widths of the core components/single Gaussian lines. This is could be caused by overestimation of the stellar velocity dispersion obtained with SDSS aperture, due to the contribution of the rotational broadening and inclination effect, as found in \cite{Eun2017}.

To  summarize, our analysis implies that non-gravitational (outflow) contribution is dominantly present in wing components of the all analysed lines, while it also partly contributes in the [O III] and H$\beta$ core components. Since the extraction of the non-gravitational contribution from the [O III] and H$\beta$ core components is complicated task, we will focus only to the wing components of all considered lines (green dots in Fig. \ref{fig2_1} and Fig. \ref{fig2_1_1}) in order to investigate the pure non-gravitational kinematics of different lines.

 \subsection{Presence of the wing components in line profiles }\label{4.0.1}

We found that the number of detected wing components among observed emission lines is significantly different. The wing components the most frequently occur in the [O III] line profiles. Moreover, in the large number of spectra, wing components are seen only in the [O III] lines, but not in the other lines (H$\beta$, H$\alpha$, [N II] and [S II]).  In these cases, [O III] wing components are weak and just slightly broader than the [O III] core component. Opposite cases, that wing components are seen in some other lines, but not in [O III] are rare. 

In total sample of 577 objects, the wing components are detected in the [O III] profile for 426 objects, in the [S II] profiles for 150 objects and in the H$\beta$ lines in only 52 objects. Similar trend is present if we consider only 'unblended' subsample (346 objects), where H$\alpha$ and [N II] are also fitted without parameter constrains. In this subsample, the wing components are detected in the [O III] profiles for 229 objects, in the H$\alpha$ profiles for 70 objects, in the [N II] profiles for 85 objects, in the [S II] profiles for 35 objects, and in the H$\beta$ profiles for only 8 objects. To understand the results of the decomposition it is very important to understand is 'missing' of the wing components in single Gaussian lines caused by physical reason or by applying the strict requirements (A$_w$/N $>$ 3 and F-test, see Section \ref{2.3.1}) for including the wing component in model of decomposition?

\begin{figure} 

\includegraphics[width=70mm]{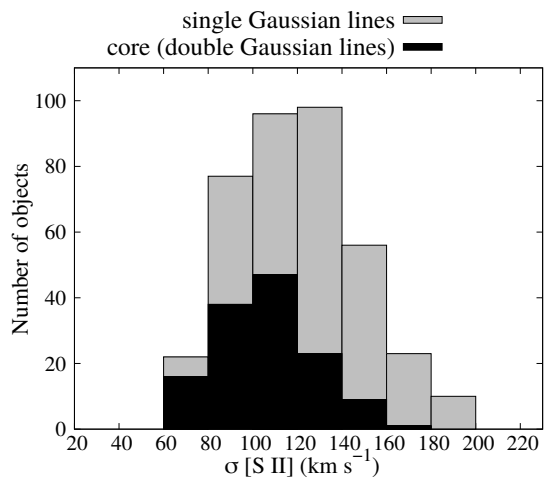}
\caption{Histogram of the width distribution of the single Gaussian [S II] lines and the core components of the double Gaussian [S II] lines.  \label{fig2_2}}
 \end{figure}

The most frequently presence of the [O III] wing components in total sample could be partly explained by the significantly stronger intensity of the [O III] lines comparing the other analysed emission lines in spectra, specially [S II] and H$\beta$, which have generally small intensities in this sample,  and similar fitting uncertainty due to noise. Although the spectra are chosen to have high S/N, it is still possible to wing components of the [S II] and H$\beta$ lines be diluted by the noise, i.e. reduced by the requested criterium of A$_w$/N $>$ 3 for the wing components.
Additionally, the weak H$\beta$ wing components could be lost in process of host galaxy subtraction. Namely, host galaxy spectra may have absorption in  H$\beta$ line, and in these cases, it is very difficult to correctly reproduce the wing component in H$\beta$ emission line profile. In 52 objects, where  H$\beta$ wing components are detected, the host galaxy spectra have no absorption in  H$\beta$, or this absorption is very small. One another concern is possibility that the wing components which are not shifted comparing the core components and which are not significantly broader than them, could be reduced by the F-test. Namely, if the line profiles could be fitted very well with single Gaussian model, including of the additional Gaussian component is not statistically justified. We found that the average widths of the single Gaussian lines are slightly broader than the average widths of the core components of double Gaussian lines. Only in the case of the [O III], these average values are similar. The histogram of distribution of the single Gaussian widths and core component widths for [S II] lines is given in Fig. \ref{fig2_2}. It could be seen that in $\sim$ 10 \% of the single Gaussian [S II] lines, the width exceed the range of the core component widths. The similar trend is observed for the H$\beta$ line. These objects are possible candidates for hidden wing components which remain undetected because of our conservative criteria for wing component detection. We can conclude that the number of detected wing components for different lines, beside physical reasons, could also be affected by the noise for the weak lines, the small shift of the wing components relative the line core and host galaxy absorption, which is specially important for the H$\beta$ lines.

\subsection{Wing component correlations in 'unblended' subsample}\label{4.1}

In this Section we used the 'unblended' subsample in which [N II] and H$\alpha$ lines are fitted independently, in order to find some relationships between wing components of analysed emission lines, which can be used for reducing the fitting parameters in case of the complex and blended
[N II]$+$H$\alpha$ wavelength band. The 'unblended' subsample contains 346 spectra, from which 229 AGN spectra have at least one wing component present.

\begin{figure}
 \centering
\includegraphics[width=80mm]{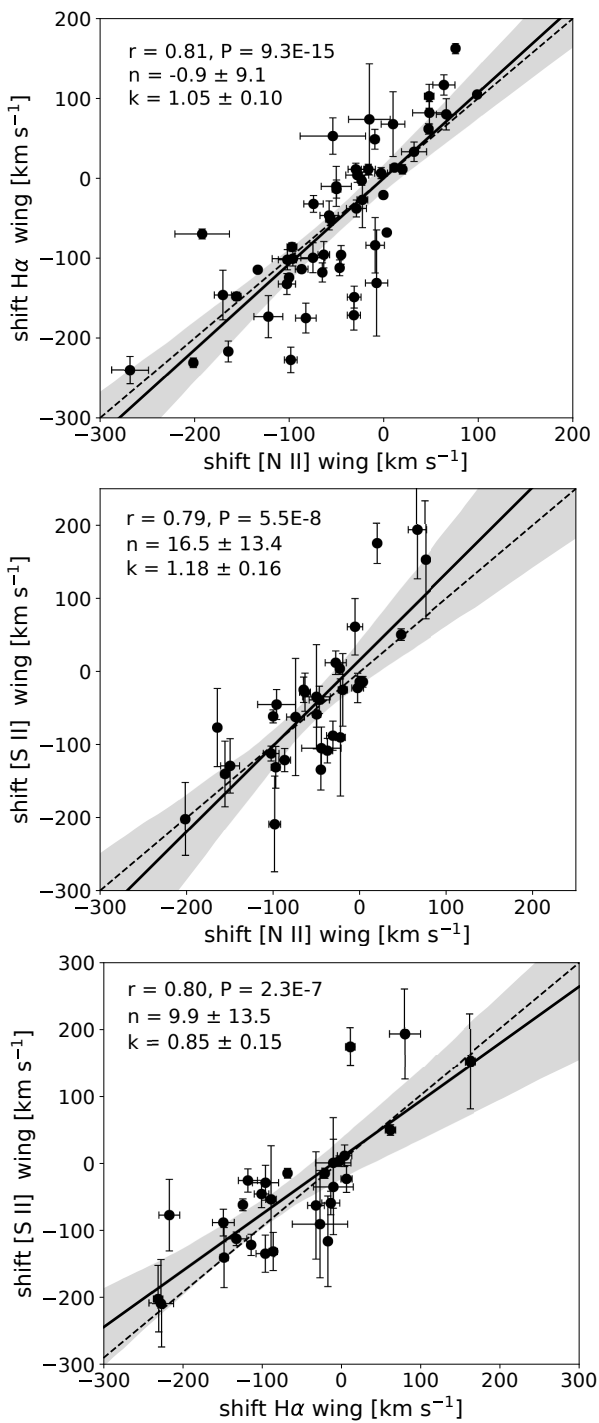}
\caption{ Correlations between the shifts of the  [N II], [S II] and H$\alpha$ wing components in 'unblended' subsample (r - coefficient of correlation, and appropriate P-value are given). The data is fitted with linear function $Y=k \cdot X+n$ (solid line), where $k$ is the slope and $n$ intercept of the best linear fit. The one-to-one relation is denoted with dashed line. The shaded region corresponds to the 95\% confidence interval. \label{fig5}}
 \end{figure}

The most significant correlations are found between the shifts of the [N II], H$\alpha$ and [S II] wing components (see Fig.~\ref{fig5}). The Pearson coefficient of the correlation between shifts of the [N II] and H$\alpha$ wings is r = 0.81, P = 9.3E-15, between [N II] and [S II] wings is r = 0.79, P = 5.5E-8 and between  H$\alpha$ and [S II] wings is r = 0.80, P = 2.3E-7, where P-value is the probability of null hypothesis that there is no correlation.
 The number of objects included in these correlations depends on the number of detected wing components in both lines. For [N II] and H$\alpha$ it is 57, for [N II] and [S II] it is 32, and for  H$\alpha$ and [S II], 30 objects (the last two have $\sim$90\% of common objects).

To investigate the relationship between the data we applied The Method of Least Squares for Simple Linear Regression Model. In Figures, we gave the obtained parameters of the fit (slope and intercept) with parameter standard errors and we shaded the 95\% confidence interval region. The linear fit between each pair of these three lines gives one-to-one relationship, within the errorbars.

\begin{figure} 
 \centering
\includegraphics[width=80mm]{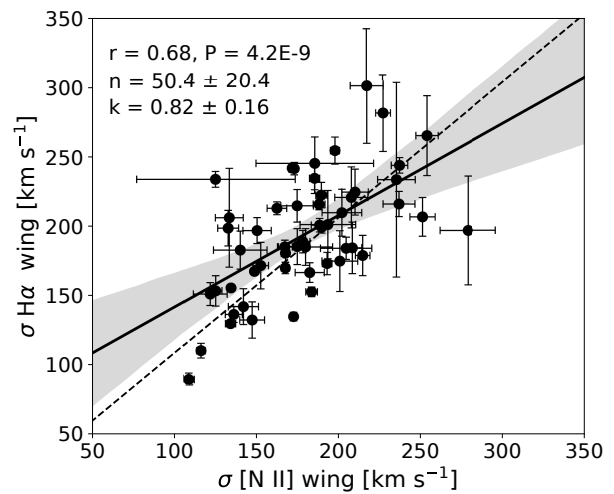}
\caption{ Correlations between the widths of the [N II] and H$\alpha$ wing components in 'unblended' subsample. Notation is the same as in Fig.~\ref{fig5}.
\label{fig7}}
 \end{figure}

Since 'unblended' subsample is chosen to [N II] and H$\alpha$ wing components do not overlap at 6575 \AA \ (or overlap very slightly), this subsample is strongly biased by spectra with small widths of the wing components. Therefore, the lack of the width-width correlations between wing components is expected because of narrow range of the wing component widths. However, despite a limited sample, the significant correlation  is found between the widths of the [N II] and H$\alpha$ wing components (r = 0.68, P = 4.2E-9), which is slightly lower than between their shifts. The linear fit also gives approximately one-to-one relationship  (see Fig.~\ref{fig7}). We noticed that, in the spectra where both, [O III] and H$\alpha$ wing components are present, the widths of the [O III] wing components are in $\sim$65\% of cases larger than those of H$\alpha$. In these spectra, the differences between the widths of the [O III] and  H$\alpha$ wing components go up to 500 km s$^{-1}$. In $\sim$35\% of cases, the widths of the H$\alpha$ wing components exceed the widths of the [O III] wing components, and their differences go up to 150  km s$^{-1}$. The similar trend is observed between the [O III] and [N II] wing component widths. These findings are in accordance with previous studies \citep{Karouzos2016a, Bae2017, Kang2017, Kang2018}, which found that the outflow contribution is generally stronger in [O III] lines comparing the H$\alpha$ lines. Therefore, we do not expect that width of the H$\alpha$ or [N II] wing components exceed much the width of the [O III] wing components in the rest of the our sample.

We found that 'unblended' subsample, as well as small subsets from 'unblended' subsample in which we found above mentioned correlations, have similar distribution of L[O III] as total sample (see Fig.2b). We assume that found correlations are maintained through the rest of the sample, which has broader wing components included. Therefore, we adopted the strongest correlations found in this subsample for fitting constrains in the rest of the spectra with blended [N II]+H$\alpha$.

\begin{figure*} 
 \centering

\includegraphics[width=150mm]{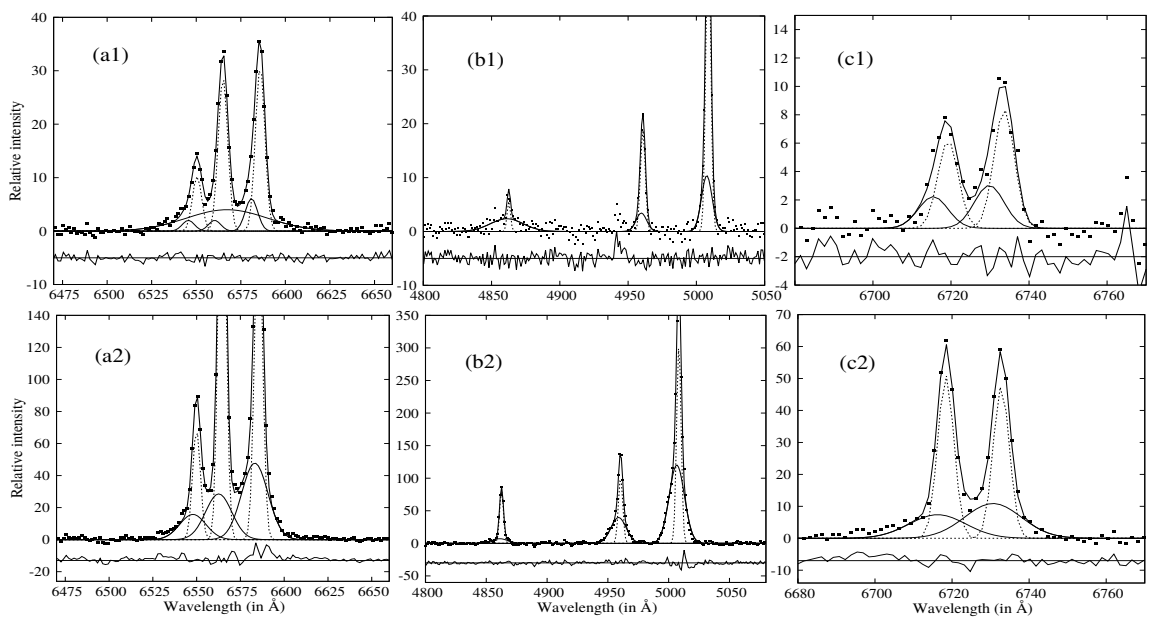} 

\caption{ Example of decomposition of objects from 'blended' subsample (group a): SDSS J010946.70+134411.9 (a1, b1, c1) and SDSS J111848.44+280738.3 (a2, b2, c2). The blended H$\alpha+$[N II]  is decomposed as: nested H$\alpha$ BLR and three weak wing components in (a1) and with three strong wing components in (a2). The decomposition of the [O III] and H$\beta$ is shown in b1 and b2, and decomposition of the [S II] lines in c1 and c2. The core components are denoted with dotted line and wing components with solid line.
\label{fig9b}}
 \end{figure*}

\subsection{Procedure for decomposition of blended [N II]$+$H$\alpha$ }\label{4.3}

 In the fitting procedure of the AGNs Type 1.8-2  spectra, the most complicated part is spectral decomposition of the blended [N II]$+$H$\alpha$ wavelength band, since in some spectra these lines have very broad wings which overlap each other. Additional complication is the possible presence of the broad H$\alpha$ component \citep[see][]{woo2014, Oh2015, Eun2017}, which arise in the BLR and in some cases it hardly differs from sum of the three broad wing components. Since in $\sim$40\% of our sample the [N II]$+$H$\alpha$ wavelength band is blended, the unraveling the complex kinematical properties in these objects is crucial for comprehensive analysis of the outflow contribution in the H$\alpha$ and [N II] lines.
 
In previous studies that have dealt with the complex [N II]$+$H$\alpha$ wavelength band, the reducing of the fitting parameters was done using the assumptions, not systematically and empirically proven relationships between fitting parameters on a large sample. The most frequently used assumptions were the same widths and shifts of the [S II] and [N II] Gaussian components, or scaling the [S II] line shape to the [N II] and H$\alpha$ lines \citep[see e.g.][etc.]{FS1988, Ho1997, popovic2004, Greene2005b, Trippe2010, Karouzos2016a, Luo2019}. Here we use the outcomes of the kinematical properties analysis of the 'unblended' subsample, presented in previous section, to give the procedure for the decomposition of the [N II]$+$H$\alpha$ wavelength band in 'blended' subsample. Presented fitting procedure can be applied only if the outflow emission in the [O III] lines can be approximated well with single Gaussian wing component. In the cases when [O III] lines have peculiar profiles with complex outflow contribution, more complex analysis is needed to decompose the [N II]$+$H$\alpha$ region (see Appendix \ref{subA}).

We noticed that spetra with  blended [N II]$+$H$\alpha$ could be divided into two groups:

\begin{enumerate} [(a)]

\item  objects with uncertain presence of the H$\alpha$ BLR in blended [N II]$+$H$\alpha$ wavelength band. In these objects, there is no flux which extend significantly out of [N II] doublet. Therefore, each of three scenarios are possible: the blended [N II]$+$H$\alpha$  can be composed of one broad H$\alpha$ component, or sum of three broaden wing components of [N II] and H$\alpha$, or in the most complicated cases, all these together --  the weak H$\alpha$ BLR nested under [N II]$+$H$\alpha$ wing components.

\item objects with certain presence of the H$\alpha$ BLR. In these objects, the broad wings of H$\alpha$ BLR component extends distinctly from both sides of the [N II] doublet, and therefore its presence is obvious even by visual inspection. 

\end{enumerate}

We established the following fitting procedure for objects  which belong to the group (a).

\begin{enumerate} 

\item Check if H$\beta$ BLR is present. In the case that H$\beta$ BLR is present in spectrum, H$\alpha$ BLR component should be included as well. H$\alpha$ and [N II] wing components should be additionally included (with reduced fitting parameters as defined below), only if obvious asymmetry is present in narrow lines.

\item If [O III] lines have no wing components detected, or if they have weak and narrow wing components (their width is not much broader than the width of the [O III] core), H$\alpha$ BLR should be included in model. In these spectra  H$\alpha$ BLR dominantly fits blended [N II]$+$H$\alpha$, and  H$\alpha$ and [N II] wing components should be included only if needed to fit the shape of the narrow lines.

\item If the [O III] lines have strong and broad wing components, then we expect that the sum of the three wing components of [N II] and H$\alpha$ dominates in blended region. Therefore, the blended [N II]$+$H$\alpha$ should be fitted with the three wing components using following fitting constrains.
 
 \begin{enumerate}[(i)]
 \item shift H$\alpha$ wing =  shift [N II] wing = shift [S II] wing. 
  \item width H$\alpha$ wing =  width [N II] wing. 
\end{enumerate}  

The [S II] lines should be fitted first, and obtained value of wing component shift should be used to fix the shift of H$\alpha$ and [N II] wing components.  If there are no [S II] wing components detected, then the shift of H$\alpha$ and [N II] wing components is one, the same, free parameter. Additionally, it is recommended to keep  H$\alpha$ and [N II] wing component widths to do not exceed much the width of the [O III] wing component (see  Sec. \ref{4.1}). In this way, we avoid possibility that three unreal, extremely broaden wing components obscure the potential presence of the one true broad H$\alpha$ component. If the fit with three wing components, limited with mentioned fitting constrains, cannot describe well the shape of the complex [N II]$+$H$\alpha$ wavelength band, then H$\alpha$  BLR component should be included. 

\end{enumerate}

The objects from the group (b), where the H$\alpha$ BLR component distinctly extends from both sides of the [N II] doublet, should be fitted  with one broad component, and  the wing components should be included only if narrow lines show asymmetry. In the case that they are needed, the wing components should be fitted with parameters constrains as defined above, for group (a). Generally, H$\alpha$  BLR component should be adopted only if it satisfies the criterium of amplitude-to-noise ratio to be larger than 3. Otherwise, it should be neglected, since it is unreliable.

\begin{figure} 
 \centering
\includegraphics[width=70mm]{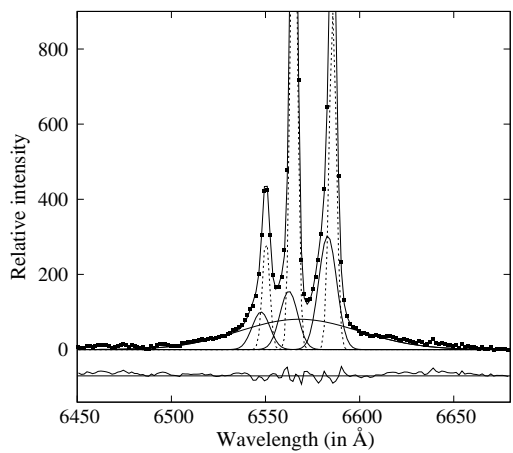}
\caption{ The example of the complex decomposition of the [N II]$+$H$\alpha$ wavelength band for object (SDSS J032525.36-060837.8) with distinctly extended H$\alpha$ BLR component (group b). The [N II]$+$H$\alpha$  is decomposed to the [N II] and H$\alpha$ core and wing components, and one  H$\alpha$ BLR component. The core components are denoted with dotted line, and wing components and H$\alpha$ BLR with solid line.
\label{fig9a}}
 \end{figure}

 By applying very careful spectral decomposition of objects from group (a), we found only 9 nested H$\alpha$ BLR components, with Full Width of the Half Maximum (FWHM) within range  2050-3000 km s$^{-1}$, while the rest of the objects from group (a) are fitted with sum of the three wing components. The examples of the spectral decomposition of the objects from the group (a), with and without nested H$\alpha$ BLR component are shown in Fig.~\ref{fig9b}.
We found 46 objects from 'blended' subsample which have obviously present very broad H$\alpha$ BLR component (group (b)). Their FWHMs are in the range of 3040-10600 km s$^{-1}$, which makes them much broader than wavelength range of [N II] doublet. The example of the fit of the complex spectra from group (b) is shown in Fig.~\ref{fig9a}.

To summarize, we found that 55 objects belong to the Type 1.9/1.8 AGNs, which makes $\sim$25\% of spectra with blended [N II]$+$H$\alpha$ in our sample. The rest of objects are Type 2 AGNs with strong wing components, which sum could be misinterpreted as broad H$\alpha$, as noticed in some other studies as well \citep{woo2014, Eun2017}.

\begin{figure} 
 \centering
 \includegraphics[width=80mm]{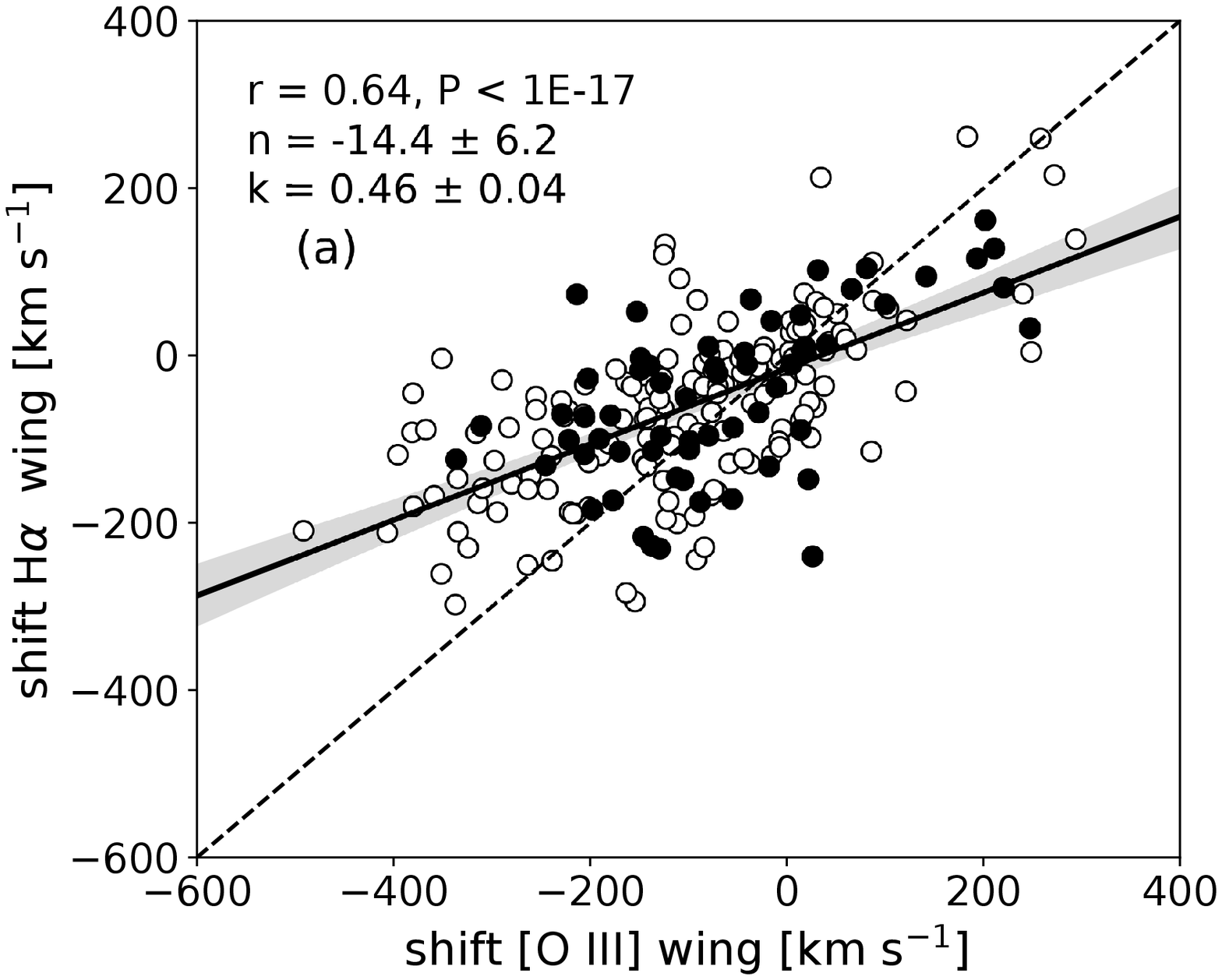} 
\includegraphics[width=80mm]{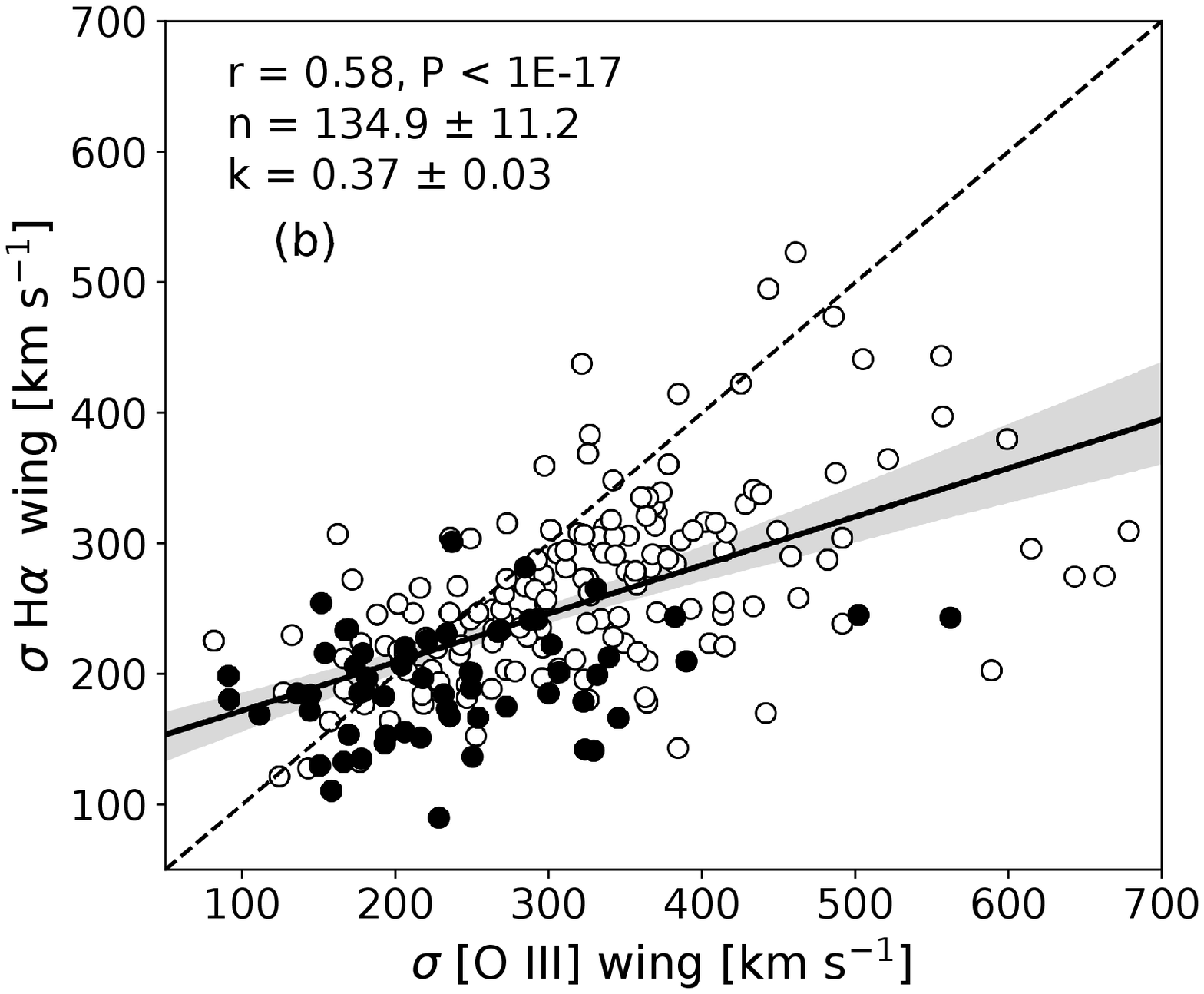}  

\caption{ Correlation between the shifts (up) and widths (bottom) of the [O III] and H$\alpha$ wing components for total sample. The 'unblended' subsample is denoted with full and 'blended' subsample with open circles. The Pearson coefficient of correlation (r), P-value, the slope ($k$) and intercept ($n$) of the best linear fit  (solid line) are given. The dashed line represents one-to-one relation and the shaded region corresponds to the 95\% confidence interval. 
\label{fig10}}
 \end{figure}

\subsection{Correlations between wing components in total sample}\label{4.5}

After we decomposed [N II]$+$H$\alpha$ in 'blended' subsample, we merged two subsamples, 'unblended' and 'blended', in order to check existing of some correlations between wing components in total sample. Of course, since we reduced fitting parameters in 'blended' subsample as described in Section \ref{4.3}, we analysed relations only between the independent line parameters for total sample. These are: width and shift of H$\alpha$ wing components, width of [S II], and widths and shifts of the H$\beta$ and [O III] wing components, which are fitted independently in total sample. 

\begin{figure} 
 \centering
 \includegraphics[width=80mm]{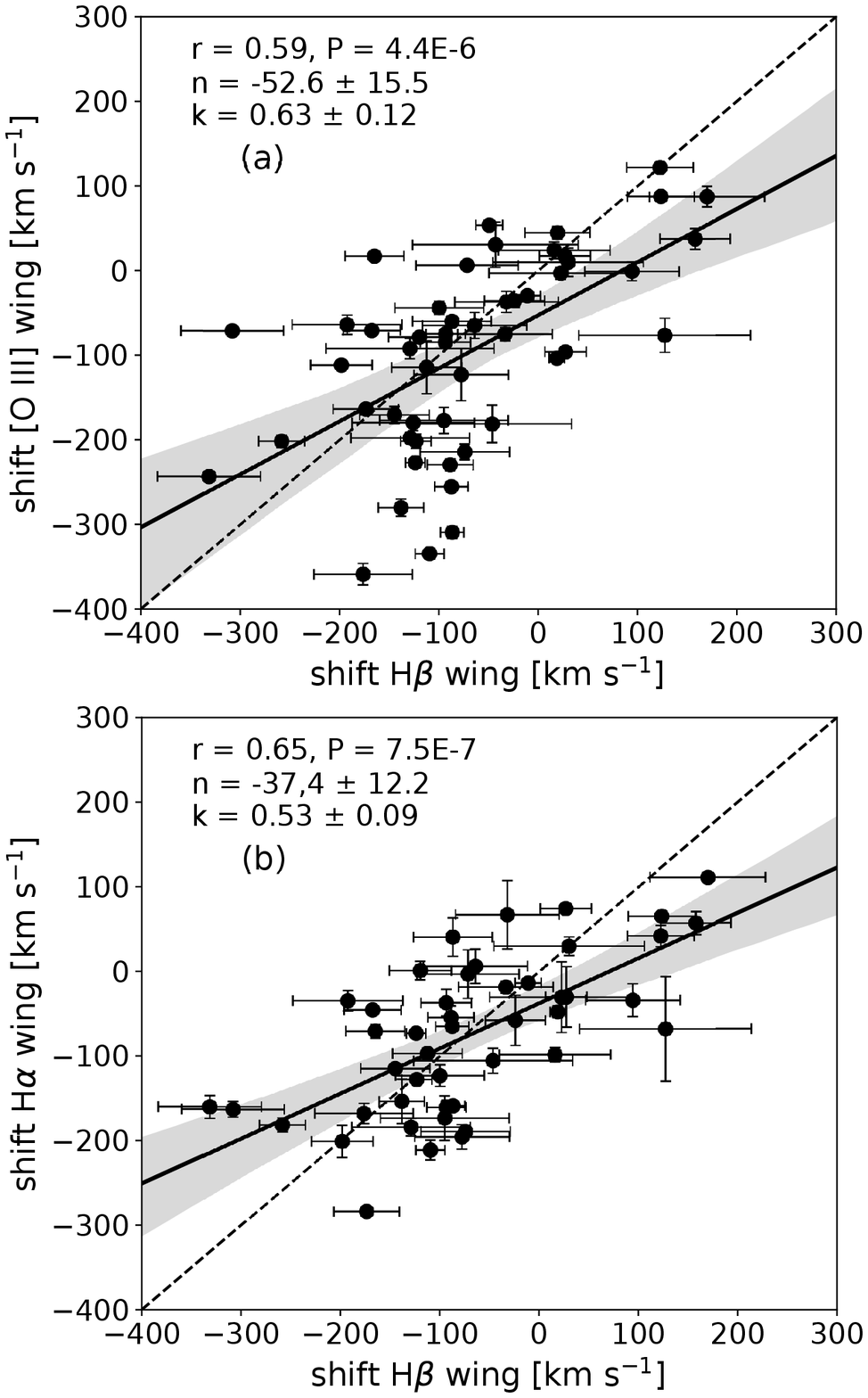} 
\caption{ Correlations between the shift of the  H$\beta$ wing components and shifts of the [O III] (up) and H$\alpha$ (down) wing components for total sample. Notation is the same as in the Fig \ref{fig5}.
\label{fig10b}}
 \end{figure}

We found that the wing component shifts of all analysed lines are in significant correlation, as well as their widths, with exception of the width of H$\beta$ wings which do not correlate with widths of the other lines. The correlations between mentioned line parameters are given in Table \ref{T2}.

The coefficient of correlation between wing component shifts of the [O III] and H$\alpha$ is r = 0.64, P < 1E-17 and this relationship is shown in Fig.~\ref{fig10}a, where 'blended' and 'unblended' subsamples are denoted with different symbols. The significant correlations between the shift of the  H$\beta$ wing components vs. shifts of the [O III] and H$\alpha$ wing components are shown in Fig.~\ref{fig10b}.

The total sample  has larger range of the wing component widths comparing the 'unblended' subsample, so we found significant correlations between the widths of the [O III], H$\alpha$ and [S II] wing components (r = 0.41 - 0.58, see Table \ref{T2}), which could not be seen in subsample with smaller range of the widths. The only exception, H$\beta$ wing components, have the weakest statistics (only 52 objects) and the least reliable fit comparing the other wing components, because  of low intensity of H$\beta$ lines and possible influence of the host galaxy absorption to the H$\beta$ wing component profiles. Also, H$\beta$ line is the one for which the physical meaning of the decomposition to the core$+$wing components is very doubtful, since H$\beta$ core components show no correlation with $\sigma_*$ (see Section \ref{4.0}).  The correlation between the widths of the  [O III] and H$\alpha$ wing components is shown in Fig.~\ref{fig10}b. Although there is large dispersion of data, the correlation is significant with coefficient of correlation r = 0.58, P < 1E-17. 

 It is specially interesting to analyse the slopes of the shift-shift and width-width correlations between the wing components of different lines. In the case of the H$\alpha$ and [N II] wing components in 'unblended' subsample, the relationships were approximately one-to-one, within uncertainties. However, in the case of the H$\alpha$ and [O III], the slopes are much shallower, even when we consider uncertainties of the linear fit. The slopes indicate generally larger widths and larger blueshifts of the [O III] wing components comparing the H$\alpha$. The difference becomes more significant with increase of the [O III] wing component blueshift (< -200 km s$^{-1}$) or  the [O III] wing component width (> 250 km s$^{-1}$), which can be seen in Fig. \ref{fig10}.

\begin{table*}
\begin{center}
\caption{Correlations between the widths and shifts of the wing components (w) of different lines in total sample ('unblended' $+$ 'blended').  \label{T2}}
\tiny

\begin{tabular}{|c c c c c c c c c|}
\hline

 & & width H$\alpha$ w & shift H$\alpha$ w &  width [S II] w &  width  H$\beta$ w  &   shift  H$\beta$ w   & shift [O III] w  &width [O III] w  \\
\hline
 \hline
    \multirow{3}{*}{  width H$\alpha$ w}&r  & 1	&0.15 &{\bf	0.53} & 0.34 	& 0.19		&0.016	& {\bf 0.58}

	\\
    &P & --	&0.02 &	{\bf 5.9E-11}&	0.02	& 0.20	&0.81&	{\bf 0}\\
&n & --	& (237) &	(143) &	(52)	& (52)	& (237) &	(237)
\\
\hline
    \multirow{3}{*}{ shift H$\alpha$ w}&r  &0.15&	1&	0.11&0.39	&{\bf	0.65}&{\bf		0.64}	&0.07

	\\
    &P & 0.02&	--&	0.21&0.007	&{\bf	7.5E-7}&	{\bf 0}&	0.28

\\
&n & (237)	& -- &	(143) &	(52)	& (52)	& (237) &	(237)
\\

 \hline
    \multirow{3}{*}{   width [S II] w}&r &{\bf  0.53}&	0.11&	1	&0.10  & -0.05	&	0.15&	{\bf 0.41}

\\
    &P &{\bf 5.9E-11} &	0.21&	--&	0.58&0.78	&	0.07	&{\bf 3.2E-7}

\\
&n & (143)	& (143) &	-- &	(52)	& (52)	& (150) &	(150)
\\
    \hline

    \multirow{3}{*} { width  H$\beta$ w} &r  &0.34&	0.39&	0.10&	1	&0.29&	0.27&	0.25

  \\
        &P & 0.02&	0.007&	0.58&	--&	0.04&	0.05&	0.07

 \\
     &n & (52)	& (52) & (52) &	--	& (52)	& (52) &	(52)
\\ 
     
\hline
    \multirow{3}{*}   { shift  H$\beta$ w } &r  &0.19&	{\bf 0.65}&	-0.05&	0.29&	1	&{\bf 0.59}&0.08

\\
                &P & 0.20&	{\bf 7.5E-7}&	0.78&	0.04&	--	&{\bf 4.4E-6}&	0.58

 \\
  &n & (52)	& (52) & (52) &	(52)	& --	& (52) &	(52)
\\ 
    \hline

  \multirow{3}{*}   { shift [O III] w  }  &r  &0.016&{\bf 	0.64}&	0.15& 0.27	&	{\bf 0.59}&	1&	-0.05

\\
                    &P  & 0.81&{\bf 	0}&	0.07&	0.05&	{\bf 4.4E-6}&	--&	0.28

\\
  &n & (237)	& (237) & (150) &	(52)	& (52)	& -- &	(426)
\\ 
  \hline 
  \multirow{3}{*}   {  width [O III] w  }   &r   & {\bf 0.58}&	0.07&{\bf 	0.41}&0.25	&0.08	&	-0.05&	1

\\
                 &P  & {\bf 0}	&0.28&	{\bf 3.2E-7}	&0.07 &0.58 &	0.28&	--

\\
 &n & (237)	& (237) & (150) &	(52)	& (52)	& (426) &	--
\\ 
\hline

\end{tabular}
\tablefoot{The Pearson coefficient of correlation (r) and P-values are given in bold-face for correlations with P$<$1E-4. The number of objects (n) with detected wing components in both lines is given in bracket.
}
\end{center}
\end{table*}

\subsection{Comparison between mean wing component profiles }\label{4.6}

\begin{figure} 
 \centering

\includegraphics[width=70mm]{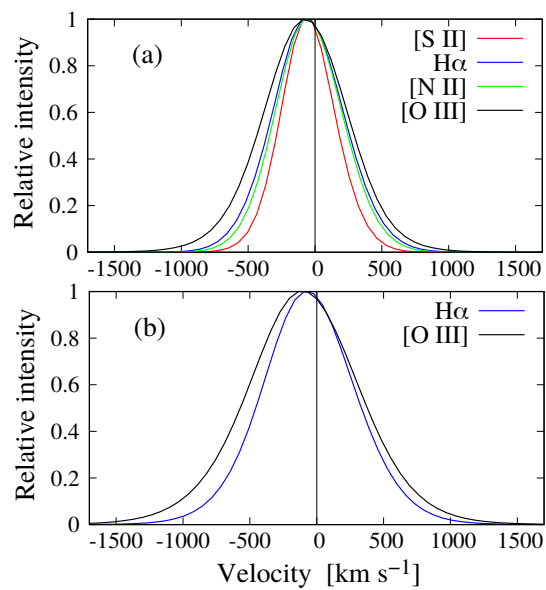}

\caption{  The mean profiles of  [O III], [N II], H$\alpha$ and [S II] wing components for 'unblended' subsample (a). The mean profiles of [O III] and H$\alpha$ wing components for 'blended' subsample (b). \label{fig13}}
 \end{figure}

The mean wing component profiles for different lines in 'unblended' subsample are shown in Fig.~\ref{fig13}a.  For making the mean profiles of the [O III], [N II] and H$\alpha$ wing components we used the spectra from 'unblended' subsample in which all these wing components are present (57 objects). The mean wing component profile of the [S II] is made by using all observed [S II] wing components in 'unblended' subsample (35 objects). The intensities of all profiles are scaled to one and shift is measured relative to the transition wavelength for each line. As it can be seen in Fig.~\ref{fig13}a, all mean profiles are slightly blueshifted relative to the their referent wavelength. The H$\alpha$ mean profile has the smallest blueshift (-36  km s$^{-1}$), while the largest blueshift is present for [O III] mean wing component profile (-90 km s$^{-1}$). The significant difference could be seen in their widths. The mean profile of the [O III] wing component is broader than the other mean profiles. The [N II] and H$\alpha$ have almost the identical mean profiles, while the narrowest is the mean [S II] profile. It seems that outflow signature in [N II] and H$\alpha$ profile is very similar.  

 In 'blended' subsample, the mean wing component profiles of [O III] and H$\alpha$ following the same trend as observed in 'unblended' subsample (see Fig.~\ref{fig13}b)\footnote{Since width and shift of [N II] wing components and shifts of [S II] wing components are fitted tied with H$\alpha$ wing components in 'blended' subsample, their mean profiles are not analysed for this subsample.}. The profiles are  made using 168 spectra, with both wing components present. As expected, both profiles have larger widths and blueshifts comparing the mean profiles of the same lines from 'unblended' subsample. Similarly as in 'unblended' subsample, the mean [O III] wing component is broader than H$\alpha$ and has slightly larger blueshift (-136 km s$^{-1}$ for [O III] and -95 km s$^{-1}$ for H$\alpha$).  The objects with extreme [O III] wing component blueshifts (< -200 km s$^{-1}$), which the most deviate from the  H$\alpha$ wing component shifts, are present in $\sim$20\% of blended spectra, so they do not strongly contribute to the shift of the mean [O III] line profile. { The broadest mean wing component profile is  seen for the [O III] lines through total sample, which indicates that outflow kinematics reflects more strongly  in  [O III] profile comparing to all other analysed lines. }

\section{Discussion}\label{5}
 Here we discuss the results presented in previous Section from different aspects. First we analysed the problem of the outflow contribution estimation in emission line profile and the complex shapes of the outflow contribution. Then we discussed the systemic influence of the outflow kinematics to the line profiles in spectra, and possible outflow emission region stratification. Finally, we draw attention to the problem of the distinction of the sum of the three wing components from the true broad H$\alpha$, in the blended [N II]$+$H$\alpha$ wavelength band.
\subsection{Extracting the outflow contribution in line profiles }\label{5.1}

The biconical outflow model predicts large diversity of complex shapes for outflow contribution, which depend on parameters as outflow velocity, inclination, amount of extinction material, etc \citep[see][]{Bae2016}. In accordance with theoretical expectations, we observed complex shapes of the [O III] (and in some cases also in H$\beta$) in  2\% of our sample of SDSS spectra. Their detailed descriptions and comparison with theoretically predicted shapes of outflow contribution are given in Appendix \ref{A}. In the rest of the  98\% of sample, all considered lines can be well fitted with single or double Gaussian models (wing$+$core components). It is possible that some physical processes in the gas outflow may have a role in producing such of Gaussian-like shape. Namely, the 'Gaussianization' of the irregular and complex shape of outflow  contribution can be caused by microturbulence, which is found to be present in AGN photoionized clouds \citep[see][]{Bottorff2000, Kraemer2007}. It is possible that during the outflow propagation it comes to mixing of the hot wind fluid and cool gas clumps at the base of the wind \citep{Westmoquette2007}, which may cause the gas turbulence. Some of the possible mechanisms could be Rayleigh-Taylor instability, for a radiatively driven cloud, or Kelvin-Helmholtz instability, for a cloud entrained in a wind \citep{Kraemer2007}. These mechanisms may cause microturbulence, which may affect the shape of the outflow component making it broader and more Gaussian-like.

Although wing$+$core model describes well the line shapes in majority of the sample, we found that physical meaning of that decomposition is 
debatable for some lines. Following the results shown in Section \ref{4.0} and additionally in Appendix \ref{B} (where influence of gravitation is analysed for H$\alpha$ and [N II] using total sample), we found that
the core component can be considered as  gravitationally dominant and wing component as outflow dominant for the  H$\alpha$, [N II] and [S II] lines, while it seems that kinematics of the [O III] and H$\beta$ lines is more complicated. It appears that the outflow contribution is significantly present in their core components as well, and it makes a large dispersion in $\sigma$ core vs. $\sigma_*$ relation. 

The possible mixture of the outflow and gravitational contribution in the [O III] core components is reported by several previous studies \citep{Karouzos2016a, woo2016, Sexton2021}. Moreover, several investigations, which predominantly analysed the [O III] lines, claim that total emission line profiles (including the core component) could be dominated by outflow kinematics \citep{Liu2013, Zakamska2014, Jarvis2019, Davies2020a}. Most of these studies performed spatially resolved observations, so it is possible that outflow kinematics contributes more in the core components when observed in smaller radius, comparing the spatially integrated spectra. However, the decomposition of the entire outflow contribution in the line profile is very complicated task, and it should be the goal of some future investigation.

\subsection{The outflow signature through AGN spectrum}\label{5.4}

In this study we performed systematic analysis of the outflow kinematics influence  to the diverse lines in AGN spectra, following the kinematical properties of the wing components. We found significant correlations between wing component shifts for all analysed lines ([O III], [N II], [S II], H$\alpha$ and H$\beta$), and  between their widths (with exception of the H$\beta$ wing components, which have the least reliable results). The significant correlations found between widths and shifts of the wing components for diverse lines ([O III], [N II], [S II], H$\alpha$ and H$\beta$) represent the signature of the outflow kinematics through AGN spectrum. 

Although the correlations between wing component widths exist between almost all lines, the significance of these correlations is sligthly lower than between their shifts, and larger dispersion of the data is present. Following the biconical model given in \cite{Bae2016}, one can expect that the velocity shift of the outflow component is probably dominantly affected by the dust extinction, and width of outflow component by bicone inclination and intrinsic outflow velocity. If we assume this model, then very good correlations between wing component shifts can be explained by the same influence of the dust extinction to the observed outflow components of all emission lines in one object. On the other hand, the width of the wing component Gaussian is more complex parameter, and interpretation of its physical meaning is more complicated. In Fig.~\ref{fig12}b, we showed the complex [O III] shape with double-peak outflow component, which probably represents superposition of emission from both, approaching and receding outflow cone, in a system without dust extinction \cite[following model of][]{Bae2016}. If one fits that outflow contribution with single wing Gaussian, the obtained result would be the Gaussian wing component with shift $\sim$ 0 km s$^{-1}$, which width would actually represent double value of projected intrinsic outflow velocity (from approaching and receding outflow cone). Therefore, there is possibility that more spectra from our sample with Gaussian wing component shift $\sim$ 0 km s$^{-1}$ represent also double-peak outflow emission, just it can not be resolved due to spectral resolution. In that case, their wing component width represent double value of projected intrinsic outflow velocity, while the  widths of the strongly shifted wing components reflect projected intrinsic outflow velocity from only one side of the outflow cone. It may be the reason for larger data dispersion and lower wing component widths correlations.

\subsubsection{The influence of the outflow to profiles of different lines}\label{new}

 Although observed correlations indicate that outflow kinematics systemically affects the line profiles in AGN spectra, it seems that it reflects with different strength in profiles of different lines. The wing components occur in the [O III] lines twice as often comparing the H$\alpha$ and [N II] lines, and three times more than in the [S II] lines. The slopes of the width-width and shift-shift relationships between the [O III] and H$\alpha$ wing components indicate significantly stronger outflow signature in the [O III] lines, specially in the case of the very blueshifted or very broad [O III] wing components. On the other hand, the wing components of the H$\alpha$ and [N II] lines have one-to-one relationship between their widths, and between their shifts, reflecting similar outflow kinematics. The comparison of the mean wing component profiles of [O III], [N II], H$\alpha$ and [S II] lines  showed small difference between their shifts, but significant difference  between their  widths (Fig.~\ref{fig13}). Considering only forbidden lines, it is interesting that the width of the mean wing component profiles decreases as the critical density and ionization potential of ion decrease. Therefore, the broadest is the [O III] mean wing component, narrower the [N II] and the narrowest the [S II] mean wing component. 
 
 The relationships between the narrow line widths and atomic characteristics (ionization potential and critical density) were investigated in several previous studies \citep[see e.g.][]{DeRobertis1984, DeRobertis1986, Ludwig2012}, but for the total narrow line profiles which are affected by the rotational, core component. In this study we observed this tendency separately for the wing components, which have pure non-gravitational origin. As already mentioned, the biconical model \citep{Bae2016} predicts that outflow velocity strongly affects the width of the outflow component, which could be explanation for the tendency observed in Fig.~\ref{fig13}. If we suppose that electron density and outflow velocity are maximal closer to the central engine, and decrease along the outflow, one can expect that the forbidden lines with higher critical density will arise in an outflow structure closer to the central engine, and will be affected by stronger outflow velocity, while the lines with smaller critical densities could not be emitted in that region. In that case, the [O III] lines would dominantly originate in outflow structure closer to the central engine, [N II] farther away, and [S II] lines would dominantly originate furthest, in outflow region with smallest density and outflow velocity. Taking into account that kinematical properties of the H$\alpha$ and [N II] wing components are very similar, and that the mean wing component profiles have very similar shapes, it is possible that H$\alpha$ and [N II] wing components are dominantly emitted in the same region in outflow structure. The analysis presented  in \cite{Davies2020a} also implies stratification in outflow emitting region: H$\alpha$ and [O III] are emitted throughout most of the ionized cloud, while  much of the [S II] line originates from mostly neutral gas, close behind the ionization front, where the electron density drops dramatically.

\subsubsection{Comparison with previous studies}\label{5.4_1}

Using the large sample of the $\sim$ 37,000 SDSS low-redshift AGNs, \cite{Kang2017} compared the shifts and widths of the [O III] and H$\alpha$ lines, calculated as the first and second moments of the total line profiles. Since the total line profiles represent the sum of the gravitational and outflow contribution, their shapes reflect the physics of both emission regions. They found that  kinematical properties of total H$\alpha$ profiles correlate with those of [O III], but that the outflow contribution  is weaker in H$\alpha$ than in [O III]. These findings are in accordance with our results obtained by comparison only wing (outflow) components of these lines.

However, the relationship between wing component widths of  H$\alpha$ and [O III] performed in this work (see Fig. \ref{fig10}b) shows shallower slope comparing the relationship between the widths of total line profiles given in \cite{Kang2017}. Our results indicating that as widths of the [O III] wing component increase, they become significantly broader comparing corresponding widths of the H$\alpha$ wing components. 

The different slope in relationships between H$\alpha$ and [O III] widths found in these two studies can be caused by several reasons considering the sample properties and method of measuring the width, since we do not include in our measurement the width of the gravitational component. The large sample in \cite{Kang2017} contains the objects with significantly lower S/N comparing the sample in this study. Since they adopt the wing components only if their A$_w$/N ratio is larger than 3, it is possible that a significant number of wing components remain undetected in noise, so they do not contribute to the total line profile. Also, \cite{Kang2017} conservatively excluded from the sample all objects which are identified as hidden AGNs Type 1 in \cite{woo2014} and \cite{Eun2017}, and additionally they excluded from the analysis all the objects which are candidates to be hidden AGN Type 1, with H$\alpha$ velocity dispersion  $>$700 km s$^{-1}$.
 These kind of objects are decomposed and included in analysis in this study.

\begin{figure*} 
 \centering
\includegraphics[width=70mm]{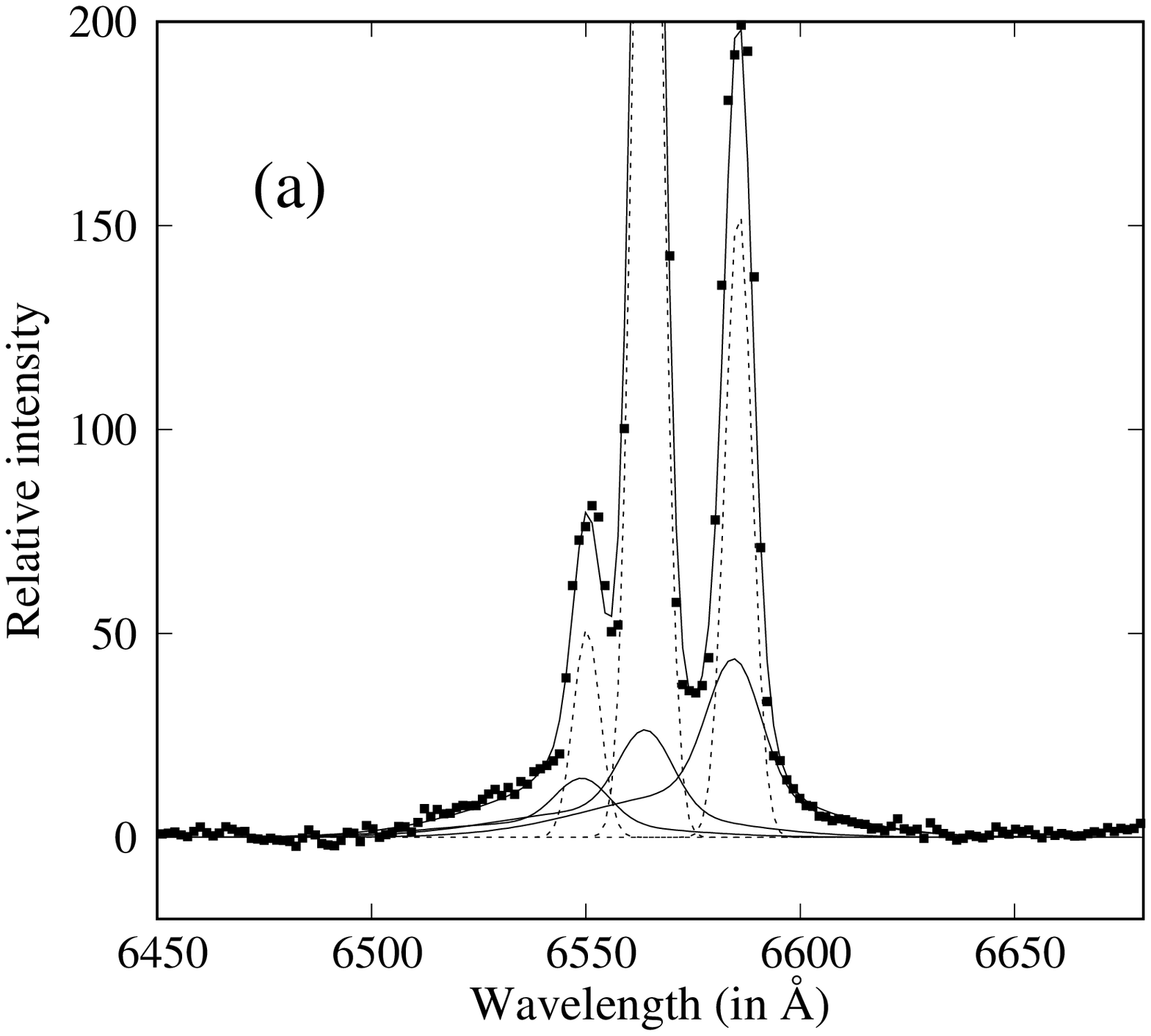}
  \includegraphics[width=70mm]{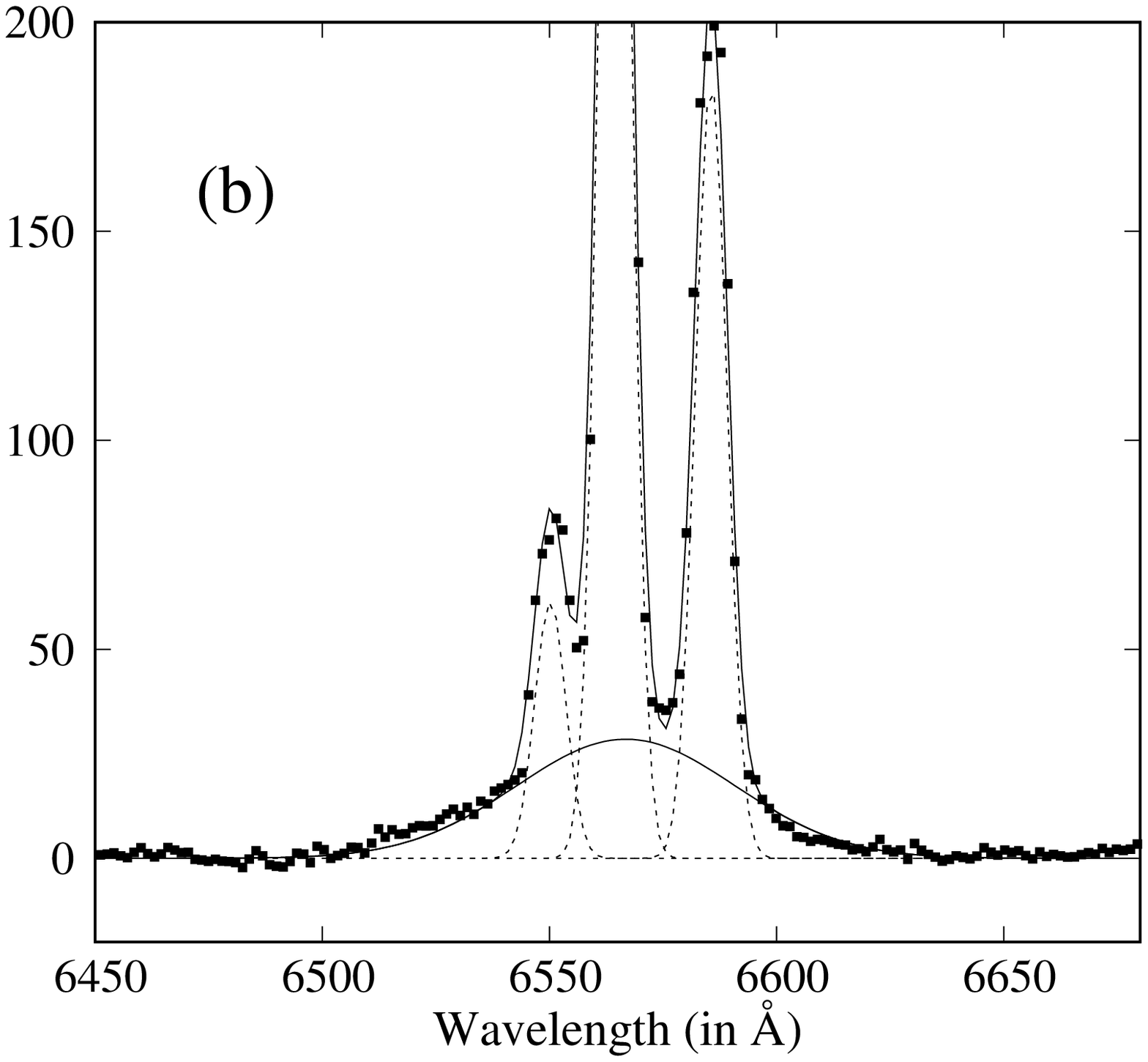}
  
\caption{ Decomposition of the  blended H$\alpha$+[N II] for object with complex shape of the [O III] outflow emission (SDSS J013706.95-090857.4, see Fig.~\ref{fig12}a). The decomposition is done with (a) sum of the three complex outflow components of H$\alpha$ and [N II] doublet, with the same shape as in [O III] lines (see Fig.~\ref{fig12}b). (b) with single broad Gaussian with FWHM$\sim$2600 km s$^{-1}$. The core components are denoted with dashed line, and outflow components/broad Gaussian with solid line.
\label{fig14}}
 \end{figure*}

\subsection{Strong outflow emission or broad H$\alpha$? }\label{5.3}

 Several studies have dealt with problem of the hidden H$\alpha$ BLR in spectra typically classified as AGNs Type 2 \citep{woo2014, Oh2015, Eun2017}. 
 One of the outcomes of the outflow kinematical analysis presented in this work is procedure for  the H$\alpha$+[N II] decomposition which should help to distinguish whether the broad H$\alpha$ is present or not in blended H$\alpha$+[N II], i.e. whether object belong to AGNs Type 2 or Type 1.9. The problem of the identification of the true H$\alpha$ BLR, as well as correct decomposition in the case of the H$\alpha$ BLR + wings superposition, is specially significant for central black hole mass (M$_{BH}$) estimation using the H$\alpha$ BLR line parameters \citep[see][]{Greene2005b}. Misinterpretation of the sum of the three broad wings of the H$\alpha$ and [N II] lines as the broad H$\alpha$ emission may lead to  wrong estimation of the M$_{BH}$ using the false H$\alpha$ BLR.

 Our results imply that multiple lines should be analysed as one system, in order to achieve physically correct spectral decomposition of the H$\alpha$+[N II] wavelength band. The misinterpretation of the H$\alpha$ BLR/wing components can be avoided by analysing the [O III] outflow contribution, which can give advantage to one model over another for H$\alpha$+[N II]  decomposition, as it is described in Section \ref{4.3}. Thus, in borderline cases, if [O III] lines have very strong and broad wing components, the blended H$\alpha$+[N II] wavelength band certainly cannot be fitted as only H$\alpha$ BLR component, with no any wing components. Also, if [O III] wing components are weak and narrow, or completely missing, we can not expect that the blended H$\alpha$+[N II] is sum of the three strong and broad wing components of H$\alpha$ and [N II]. Therefore, the shape of the [O III] outflow contribution (complex or Gaussian-like), and its strength (weak and narrow or strong and broad)  have important role in understanding the origin of the flux in blended H$\alpha$+[N II] in each spectrum. On the other hand, [S II] has two order of magnitude lower critical density than [O III] lines, so it will fade away in high density outflow region, which still could be the region of efficient emission for [O III] (also for H$\alpha$ and [N II]). Therefore, the total shape of the [S II] lines (core + wing) is not a suitable template for total shape of H$\alpha$ and [N II] narrow lines in process of decomposition of blended H$\alpha$+[N II]. Namely, the presence of the wing components in [S II] lines implies the presence of the wing components in H$\alpha$ and [N II] with the same shift (see Fig. \ref{fig5}) and generally larger widths, but their absence cannot be used as a confirmation of the non-existing of the H$\alpha$ and [N II] wing components (see Sec. \ref{4.3}).

In the case of the complex shapes of the [O III] outflow contribution (see Appendix \ref{A}, Fig.~\ref{fig12}), the decomposition of the blended H$\alpha$+[N II] is even more  complicated, and becomes a real challenge. It is expected that complex, peculiar shape of outflow emission seen in [O III] could be seen in the shapes of the other lines in spectra as well. However, it should be taken into account that  H$\beta$ and [S II] lines are weaker than [O III], H$\alpha$ and [N II] in spectra of AGNs Type 2, and therefore their complex outflow shapes can be diluted by noise,  or reduced by high density of outflow region in the case of the [S II].

To  illustrate the decomposition of H$\alpha$+[N II] in objects with strong and complex outflow emission in [O III], we used the object SDSS J013706.95-090857.4, which is analysed in Appendix \ref{A}, and shown in Fig.~\ref{fig12}a as en example of the complex outflow emission. 
We tried to decompose blended H$\alpha$+[N II] in this object by adopting the same shape of the outflow component as extracted from [O III] (see Fig.~\ref{fig12}a1), for outflow components of H$\alpha$ and [N II] lines. The details of the fitting procedure of the H$\alpha$+[N II] using this complex outflow shape are given in Appendix \ref{subA}.
 The decomposition of the blended H$\alpha$+[N II] is shown in Fig.~\ref{fig14}a. As it can be seen, the sum of the three complex outflow components, with shape as seen in [O III] lines, fits very well the emission flux under H$\alpha$+[N II]. 
 It should be pointed out that the flux of the blended H$\alpha$+[N II] in SDSS J013706.95-090857.4 distinctly extends blueward the [N II] doublet, visually making an impression that it is obviously H$\alpha$ BLR component. Without considering the extremely strong outflow emission in [O III], it could be fitted with single broad Gaussian with FWHM = 2600 km s$^{-1}$ (see Fig.~\ref{fig14}b), interpreted as H$\alpha$ BLR component, and potentially used for M$_{BH}$ estimation as it is done for this object in several studies which analyse a large data samples \citep{Greene2007, Oh2015, Liu2019}. More detailed investigation is needed to interpret the real nature of  blended H$\alpha$+[N II] in this object.

\section{Conclusions}\label{6}

In this research, we used 577 AGN Type 1.8-2 spectra obtained from SDSS, with high S/N, in order to trace the outflow kinematics. We applied very careful spectral decomposition of the several emission lines (H$\beta$, [O III], H$\alpha$, [N II], [S II]) using single and double Gaussian model (wing$+$core components), where wing components are taken to be a proxy of the outflow contribution. In order to investigate the outflow contribution in H$\alpha$ and [N II] lines in total sample, we specially focused to the unravelling of the complex kinematics of the blended [N II]$+$ H$\alpha$, which is present in $\sim$40\% of our sample. We used the outcomes of the analysis of the subsample with unblended H$\alpha$ and [N II] lines in order to obtain fitting constrains and establish the procedure for decomposition of blended [N II]$+$ H$\alpha$.

 The influence of the outflow kinematics to the line profiles is investigated by performing correlations between several kinematical parameters  and by comparing the wing component mean profiles for different lines, using the subsets in which wing components are present in analysed lines. In this way we investigated the systemic  influence of the outflow kinematics to the different lines in spectra, as well as the difference in how much is each line affected by outflow kinematics. 
 
Summarizing the obtained results, we can outline following conclusions:

\begin{enumerate} [(i)]

\item The outflow kinematics has systemic influence to the considered emission lines in AGN spectra (H$\beta$, [O III], H$\alpha$, [N II], [S II]), which can be seen through strong correlations between their wing component shifts, as well as between their wing component widths, found between each pair of lines, in subsets where wing components are detected in both compared lines. The only exception are H$\beta$ wing component widths (H$\beta$ wing components are the least reliable and have the weakest statistic). Generally, the correlations between wing component shifts are slightly stronger comparing the correlations between wing component widths for all lines. It is possible that  wing component shifts dominantly reflect the amount of the extinction material, as predicted by biconical model, while the wing component widths seems to be more complex parameters, affected with different projections of the outflow velocity, which probably cause the larger dispersion in the width-width correlations.

\item The signature of the outflow kinematics differs for different lines. By comparing  the line component dispersions with stellar velocity dispersion, we found that the wing components of all considered emission lines have pure non-gravitational kinematics. The core components  are consistent with gravitational kinematics for  H$\alpha$, [N II] and [S II] lines, while in [O III] there is evidence for contribution from non-gravitational kinematics. On the other hand, we found that all single Gaussian lines (with no wing component detected) represent  gravitationally dominant kinematics. The comparison of the mean width of the wing components in subsets in which wing components are detected in all lines, revealed that the [O III] wing components are generally broader than wing components of the other analysed lines. The widths of the H$\alpha$ and [N II] wing components are very similar, while the widths of the [S II] wing components are generally the narrowest. If we assume that outflow velocity strongly affects the width of outflow component, as predicted by biconical model, then the difference in widths may imply that these lines dominantly arise in different parts of an outflowing region.

\item  The blended H$\alpha+$[N II], which can be the sum of the three strong wing components, the true broad H$\alpha$, or all these together, can be successfully decomposed using the outcomes of the analysis of the subsample with unblended H$\alpha$ and [N II] lines. Obtained parameter constrains prevent misinterpretation of the sum of three wing components of H$\alpha$ and [N II] as H$\alpha$ BLR or vice versa. Generally, in order to achieve physically correct spectral decomposition of blended H$\alpha+$[N II], multiple lines in spectra should be analysed as one system, since their shapes systemically reflect the physical conditions in outflow emission region. One should be especially careful if extended and complex emission is observed in [O III]. In that case, the sum of the complex outflow emission of  H$\alpha$ and [N II] can mimic the broad H$\alpha$ with FWHM up to $\sim$2600 km s$^{-1}$, which makes the spectral decomposition these objects very difficult.

\item Although the biconical outflow model predicts large diversity of complex shapes for outflow contribution, we found the complex [O III] shapes only in $\sim$2\% of the sample, while the rest of the sample can be well fitted with single or double Gaussian models. It is possible that microturbulence, which could arise due mixing of the hot wind fluid and cool gas during the outflow propagation, is responsible for making the complex outflow line shapes more 'Gaussian-like' in majority of the spectra.

\end{enumerate}

\begin{acknowledgements}

 We thank anonymous referee for valuable comments and suggestions 
 that helped us to significantly improve the
paper. This work is supported by the Ministry of Education, Science and Technological Development of Serbia (451-03-9/2021-14/20016 and 451-03-68/2020-14/200002).

Funding for the Sloan Digital Sky 
Survey IV has been provided by the 
Alfred P. Sloan Foundation, the U.S. 
Department of Energy Office of 
Science, and the Participating 
Institutions. 

SDSS-IV acknowledges support and 
resources from the Center for High 
Performance Computing  at the 
University of Utah. The SDSS 
website is www.sdss.org.

SDSS-IV is managed by the 
Astrophysical Research Consortium 
for the Participating Institutions 
of the SDSS Collaboration including 
the Brazilian Participation Group, 
the Carnegie Institution for Science, 
Carnegie Mellon University, Center for 
Astrophysics | Harvard \& 
Smithsonian, the Chilean Participation 
Group, the French Participation Group, 
Instituto de Astrof\'isica de 
Canarias, The Johns Hopkins 
University, Kavli Institute for the 
Physics and Mathematics of the 
Universe (IPMU) / University of 
Tokyo, the Korean Participation Group, 
Lawrence Berkeley National Laboratory, 
Leibniz Institut f\"ur Astrophysik 
Potsdam (AIP),  Max-Planck-Institut 
f\"ur Astronomie (MPIA Heidelberg), 
Max-Planck-Institut f\"ur 
Astrophysik (MPA Garching), 
Max-Planck-Institut f\"ur 
Extraterrestrische Physik (MPE), 
National Astronomical Observatories of 
China, New Mexico State University, 
New York University, University of 
Notre Dame, Observat\'ario 
Nacional / MCTI, The Ohio State 
University, Pennsylvania State 
University, Shanghai 
Astronomical Observatory, United 
Kingdom Participation Group, 
Universidad Nacional Aut\'onoma 
de M\'exico, University of Arizona, 
University of Colorado Boulder, 
University of Oxford, University of 
Portsmouth, University of Utah, 
University of Virginia, University 
of Washington, University of 
Wisconsin, Vanderbilt University, 
and Yale University.
\end{acknowledgements}



\begin{appendix}

\section{The complex shapes of the [O III] outflow components and biconical model}\label{A}

The complex shapes of the  [O III] lines were reported by many previous studies \citep[see e.g.][]{Crenshaw2010, Harrison2014, Zakamska2014, Davies2020a}. Here we discuss some of the complex shapes of the [O III] lines found in this paper.

\begin{figure} 
 \centering

\includegraphics[width=90mm]{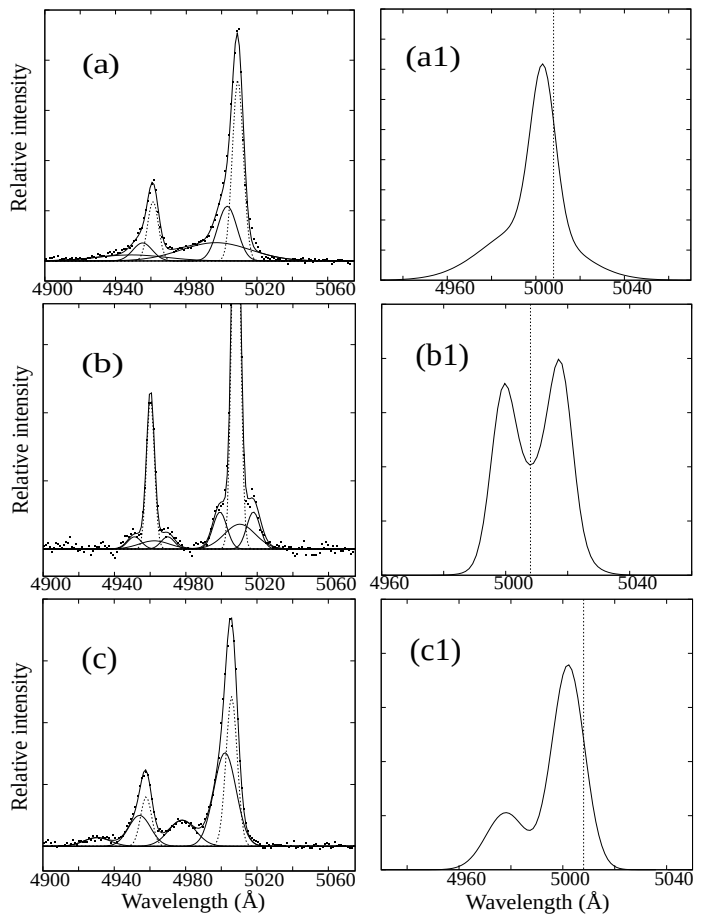}

\caption{ Three specific examples of the complex [O III] shapes seen in (a) SDSS J013706.95-090857.4, (b) SDSS J163453.66+231242.6 and (c) SDSS J165121.88+215526.2 objects. The complex shape decomposition is shown in left side of Figures, where the core components are denoted with dashed line, and multiple Gaussians which sum gives the outflow component, with solid line. The outflow components for [O III]$\lambda$5007 \AA \ are shown separately, in right side of Figures (a1, b1, c1), where vertical dashed line represents zero velocity shift. 
\label{fig12}}
 \end{figure}

In Fig.~\ref{fig12} we have singled out three specific examples of the [O III]  complex shapes from our sample. The majority of the complex  [O III] profiles  have the shape similar as in Fig.~\ref{fig12}a, while the [O III] shapes shown in Fig.~\ref{fig12}b  and Fig.~\ref{fig12}c are seen only in these two spectra. The [O III] profiles are fitted with one narrow, core component, and the outflow component is fitted with two or three Gaussians. These Gaussians are used only to approximately describe the complex shapes of outflow emission in these specific spectra, without presumption of physical meaning for each Gaussian individually. Similar as for the simple cases, the widths and shifts of outflow component Gaussians are the same for both [O III] doublet lines, but intensity ratio is fixed to be $\sim$ 3. The sum of these Gaussians, which represents the outflow contribution,  is shown separately in right side of Fig.~\ref{fig12} for [O III]$\lambda$ 5007 \AA \ line.

\cite{Bae2016}  gave the 3D models of biconical outflows combined with a thin dust plane. Using a set of input parameters, they found that intrinsic outflow velocity, bicone inclination, and the amount of dust extinction, mainly determine the simulated velocity and velocity dispersion of the outflow components. Also, they simulated the emission line profiles for various combinations of the input parameters, giving the diversity of the possible outflow emission shapes.

 The observed complex profile of the [O III] outflow component shown in Fig.~\ref{fig12}a is very similar to simulated line profiles given by biconical outflow model for included effects of the dust extinction and angle of bicone inclination $i= \pm 30^\circ$ \citep[see Fig.~9 in][]{Bae2016}. In this case, model predicts that dust obscures a part of the bicone, so the  emission line flux arising from the outflow offsets from zero velocity, showing asymmetric profiles. Interestingly, the complex line profile shown in Fig.~\ref{fig12}b, with double-peak outflow emission is also predicted by biconical model for the case with no dust extinction and with the outer opening angle of the bicone smaller than absolute value of the angle of bicone inclination, $i$ \citep[see Fig.~11 in][]{Bae2016}. However, the complex line profile shown in Fig.~\ref{fig12}c, with outflow emission composed from two components, with smaller one extremely blueshifted, is not presented in simulated spectra. The blueshift of the higher Gaussian  is -370 km s$^{-1}$, while smaller component is blueshifted -1810 km s$^{-1}$. It is possible that this complex outflow shape represent the projection of the two independent, spatially separated outflows.

\subsection{The decomposition of the [N II]$+$ H$\alpha$ with complex outflow shape}\label{subA}

Here we give the example of possible decomposition of the blended [N II]$+$ H$\alpha$ for object with complex outflow shape observed in [O III] lines. We chose the object SDSS J013706.95-090857.4 (shown in Fig.~\ref{fig12}a), because it is frequently analysed in literature, and it has complex outflow shape which is the most common in this sample.

The complex shape of outflow emission in [O III] for this object is approximately described as sum of the two Gaussians (see Fig.~\ref{fig12}a1), so we adopted that shape for outflow components of H$\alpha$ and [N II] lines. We found intensity ratio of these two Gaussians within one [O III] line, and also the difference between their shifts and widths. In this way we describe complex shape shown in Fig.~\ref{fig12}a1 with only three free parameters: width, shift and intensity of the higher Gaussian, while smaller Gaussian is reproduced relative to higher one. Similarly as done in fitting procedure described in Section \ref{4.3}, we fixed  H$\alpha$ and [N II] outflow components to have the same shift and width, but different intensities. Finally, we fitted blended H$\alpha$+[N II] in this object with three complex outflow components, described with only 4 free parameters (width, shift and intensity of complex outflow shape for H$\alpha$ and one more intensity for [N II] doublet). 

We found that the complex outflow components which give the best fit of the blended H$\alpha$+[N II] have 14\% smaller width and significantly smaller blueshift comparing the complex outflow component of [O III] (shift of [O III] outflow component peak $\sim$ -300 km s$^{-1}$, of H$\alpha$ and [N II] $\sim$ -30 km s$^{-1}$), which is in agreement with relationships obtained by analysing the simple Gaussian outflow components (see Fig.~\ref{fig10}). The H$\beta$ and [S II] line profiles in this objects show only small blue asymmetry. It is possible that smaller, extended and extremely broad component of the outflow complex shape cannot be resolved for these lines, because it is diluted by the noise. The decomposition of the blended H$\alpha$+[N II] is shown in Fig.~\ref{fig14}a.

\section{The correlations between H$\alpha$ and [N II] line components and $\sigma_*$ for total sample }\label{B}

In Section \ref{4.0}  we analysed the influence of the gravitation to the kinematical properties of the different line components for double Gaussian and single Gaussian H$\alpha$ and [N II] lines (see Fig. \ref{fig2_1_1}, Table \ref{T3}). The influence of gravitation to the line shape of these two lines was done only for 'unblended' subsample in which  H$\alpha$ and [N II] lines were fitted completely independently. 
 However, there is concern that these results could be biased by selection effect, since the objects with the broadest wing components in H$\alpha$ and [N II] lines are part of the 'blended' subsample. The 'blended' subsample is decomposed using some fitting constrains for wing components (the same widths and shifts for H$\alpha$ and [N II] wing components, see Section \ref{4.3}), while the core components of both lines are fitted with all free parameters. Some spectra from 'blended' subsample are decomposed to broad H$\alpha$ and single Gaussian  H$\alpha$ and [N II] lines, which are also fitted with all free parameters.
 
\begin{figure} 
 \centering

\includegraphics[width=65mm]{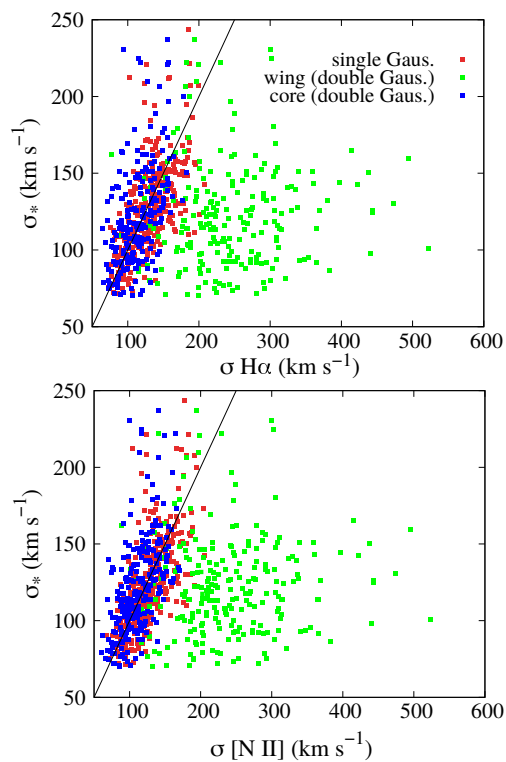}
\caption{ Correlations between the $\sigma_*$ and velocity dispersion of the single Gaussian lines (red dots), wing and core components of the double Gaussian lines (green and blue dots) for H$\alpha$ and [N II] lines (total sample). The one-to-one relation is denoted with solid line.\label{figA0}}
 \end{figure}

 Here we investigated  the  correlations between $\sigma_*$ and H$\alpha$/[N II] line components using the total sample ('unblended' + 'blended' subsample). Similarly as for 'unblended' subsample, we found no correlations between the widths of the H$\alpha$ and [N II] wing components and $\sigma_*$, and we found significant correlations between widths of the core components (and single Gaussian lines) and $\sigma_*$. The  coefficient of correlation between $\sigma_*$ and the H$\alpha$ core components for total sample is r = 0.50, P = 1.8E-15, and between $\sigma_*$ and  [N II] core components, r = 0.53, P = 0. The coefficients of correlations between $\sigma_*$ and single Gaussian lines are: r = 0.63, P = 0 for H$\alpha$, and  r = 0.64, P = 0 for [N II]. The relationships are shown in Figure \ref{figA0}.
 
 We can conclude that including the 'blended' subsample in analysis does not change the basic conclusions about gravitational influence to the  H$\alpha$ and [N II] line components found for 'unblended' subsample in Section \ref{4.0}.

\end{appendix}

\begin{thebibliography}{}
\bibitem[Abolfathi et al.(2018)]{Abolfathi2018} Abolfathi, B., Aguado, D. S., Aguilar, G., Allende Prieto, C. et al., 2018, \apjs, 235, 42A 

\bibitem[Bae \& Woo (2014)]{Bae2014} Bae, H.-J., Woo, J.-H. 2014, \apj, 795, 30
\bibitem[Bae \& Woo (2016)]{Bae2016} Bae, H.-J., Woo, J.-H. 2016, \apj, 828, 97
\bibitem[Bae et al. (2017)]{Bae2017} Bae, H.-J., Woo, J.-H., Karouzos, M., Gallo, E., Flohic, H., Shen, Y., Yoon, S.-Y 2017, \apj, 837, 91


 \bibitem[Baldwin et al.(1981)]{Baldwin81} Baldwin, J.~A., Phillips, M.~M., \& Terlevich, R.\ 1981, \pasp, 93, 5
 
  \bibitem[Berton \& J\"{a}rvel\"{a} (2021)]{Berton2021} Berton, M. \& J\"{a}rvel\"{a}, E., 2021, Universe, 7, 188
 \bibitem[Blanton et al. (2017)]{Blanton2017} Blanton, M.R.,  Bershady, M. A., Abolfathi, B., Albareti, F. D. et al. 2017, AJ, 154, 26B.
 
  \bibitem[Bolton et al. (2012)]{Bolton2012} Bolton, A.S., Schlegel, D.J., Aubourg, \'E., Bailey, S. et al. 2012, AJ, 144, 144.
 
  \bibitem[Bottorff et al. (2000)]{Bottorff2000} Bottorff, M., Ferland, G., Baldwin, J., Korista, K. 2000, \apj, 542, 644
 
 \bibitem[Cappellari \& Emsellem (2004)]{Cappellari2004} Cappellari M., Emsellem E. 2004, PASP, 116, 138

 
 
 \bibitem[Costa et al. (2020)]{Costa2020} Costa, T.,  Pakmor, R., Springel, V. 2000, \mnras, 497, 5229
 
  \bibitem[Crenshaw et al. (2010)]{Crenshaw2010} Crenshaw, D. M.,  Schmitt,  H. R., Kraemer, S. B., Mushotzky, R. F.,  Dunn, J. P. 2010, \apj, 708, 419
  
 \bibitem[Davies et al. (2020a)]{Davies2020a} Davies, R., Baron,  D. , Shimizu, T., Netzer, H. et al. 2020a, \mnras, 498, 4150
\bibitem[Davies et al. (2020b)]{Davies2020b}  Davies, R., F\" {o}rster Schreiber, N. M., Lutz, D. et al.  2020b, \apj, 894, 28
  
  \bibitem[Dawson et al. (2013)]{Dawson2013}  Dawson, K.S., Schlegel, D.J., Ahn, C.P., Anderson, S.F. et al.  2013, \aj, 145, 10
  
  
  \bibitem[De Robertis  \& Osterbrock (1984)]{DeRobertis1984}  De Robertis, M. M., \& Osterbrock, D. E. 1984, \apj, 286, 171
  \bibitem[De Robertis  \& Osterbrock (1986)]{DeRobertis1986}  De Robertis, M. M., \& Osterbrock, D. E. 1986, \apj, 301, 727
 
 \bibitem[Di Matteo et al. (2005)]{DiMatteo2005} Di Matteo, T., Springel, V., Hernquist, L. 2005, Nature, 433, 604
 
 \bibitem[Dimitrijevi\'c et al.(2007)]{dim07} Dimitrijevi\'c, M.~S., Popovi\'{c}, L.~\v{C}., Kova\v{c}evi\'{c}, J., Da\v{c}i\'{c}, M., Ili\'{c}, D.\ 2007, \mnras, 374, 1181

 \bibitem[Doj\v cinovi\'c et al. (2022)]{Dojcinovic2022} Doj\v cinovi\'c, I., Kova\v{c}evi\'{c}-Doj\v cinovi\'c, J., Popovi\'{c}, L.~\v C., 2022, submitted in ASR

\bibitem[Eun et al. (2017)]{Eun2017} Eun, D., Woo, J.-H. \&  Bae, H.-J. 2017,  \apj, 842, 5

\bibitem[Everett (2005)]{Everett2005} Everett, J. E. 2005,  \apj,  631, 689

\bibitem[Fabian et al. (2012)]{Fabian2012} Fabian, A. C. 2012,  ARA\&A, 50, 455
 \bibitem[Filippenko \& Sargent (1988)]{FS1988} Filippenko, A.V., Sargent, W.L.W. 1988, \apj, 324, 134
 
 \bibitem[Greene \& Ho (2005a)]{Greene2005a} Greene, J. E. \& Ho, L. C. \ 2005a, \apj, 627, 721

\bibitem[Greene \& Ho (2005b)]{Greene2005b} Greene, J. E. \& Ho, L. C. \ 2005b, \apj, 630, 122

\bibitem[Greene \& Ho (2007)]{Greene2007} Greene, J. E. \& Ho, L. C. \ 2007, \apj, 667, 131
 
 \bibitem[Harrison et al. (2014)]{Harrison2014} Harrison, C. M., Alexander, D. M., Mullaney, J. R., Swinbank, A. M. 2014, \mnras, 441, 3306
\bibitem[Harrison et al. (2018)]{Harrison2018} Harrison, C. M., Costa, T., Tadhunter, C. N., Flütsch, A., Kakkad, D., Perna, M., Vietri, G. 2018, Nature Astronomy, 2, 198
 
 
 \bibitem[Ho et al. (1997)]{Ho1997} Ho, L. C., Filippenko, A. V., Sargent, W. L. W., Peng, C. Y. 1997, \apjs, 112, 391

\bibitem[Howarth (1983)]{ho1983} Howarth, I.~D.\ 1983, \mnras, 203, 301

\bibitem[Ishibashi \& Fabian (2015)]{Ishibashi2015} Ishibashi, W. \&  Fabian A. C. 2015, \mnras, 451, 93



\bibitem[Jarvis et al. (2019)]{Jarvis2019} Jarvis, M. E., Harrison, C. M., Thomson, A. P., Circosta, C. et al. 2019, \mnras, 485, 2710
\bibitem[Jarvis et al. (2021)]{Jarvis2021} Jarvis,  M. E., Harrison, C. M., Mainieri, V., Alexander, D. M. et al. 2021, \mnras, 503, 1780

\bibitem[Kang et al. (2017)]{Kang2017} Kang, D., Woo, J.-H.  \&  Bae, H.-J. 2017, \apj, 845, 131
\bibitem[Kang \& Woo (2018)]{Kang2018} Kang, D. \& Woo, J.-H. 2018, \apj, 864, 124

\bibitem[Karouzos et al. (2016a)]{Karouzos2016a} Karouzos, M., Woo, J.-W. and Bae, H.-J. 2016a, \apj, 819, 148
\bibitem[Karouzos et al. (2016b)]{Karouzos2016b} Karouzos, M., Woo, J.-W. and Bae, H.-J. 2016b, \apj, 833, 171






\bibitem[Komossa et al. (2018)]{Komossa2018} Komossa, S., Xu, D., Wagner, A.Y. 2018, \mnras,  477, 5115


\bibitem[Kova\v{c}evi\'{c} et al.(2010)]{kovacevic2010} Kova\v{c}evi\'{c}, J., Popovi\'{c}, L.~\v{C}. \& Dimitrijevi\'{c}, M.,\ 2010, \apjs, 189, 15
\bibitem[Kraemer et al. (2007)]{Kraemer2007} Kraemer, S. B., Bottorff, M. C.  \&  Crenshaw, D. M. 2007, \apj,  668, 730



\bibitem[Kuraszkiewicz et al. (2002)]{ku2002} Kuraszkiewicz, J.~K., Green, P.~J., Forster, K., Aldcroft, T.~L., Evans, I.~N., Koratkar, A. 2002, \apjs, 143, 257


\bibitem[Liu et al.(2019)]{Liu2019} Liu, H.-Y., Liu, W.-J., Dong, X.-B., Zhou, H., Wang, T., Lu, H., Yuan, W. 2019, \apjs, 243, 21

\bibitem[Liu et al.(2013)]{Liu2013} Liu, G., Zakamska, N. L., Greene, J. E., Nesvadba, N. P. H., Liu, X. 2013, \mnras, 436, 2576


\bibitem[Ludwig et al. (2012)]{Ludwig2012} Ludwig, R. R., Greene, J. E., Barth, A. J., Ho, L. C. 2012, \apj,  756, 51

\bibitem[Luo et al. (2019)]{Luo2019} Luo, R., Woo, J.-H., Shin, J., Kang, D., Bae, H.-J., Karouzos, M. 2019, \apj,  874, 99

\bibitem[Lupton (1993)]{Lupton1993} Lupton, R. H.  1993, Statistics in Theory and Practice (Princeton: Princeton
Univ. Press)


\bibitem[Molyneux et al. (2019)]{Molyneux2019} Molyneux, S. J., Harrison, C. M. \& Jarvis, M. E. 2019, A\&A, 631, 132


\bibitem[Mukherjee et al. (2018)]{Mukherjee2018} Mukherjee, D., Bicknell, G. V., Wagner, A. Y., Sutherland, R. S. \& Silk, J. 2018, \mnras, 479, 5544

\bibitem[Mullaney et al. (2013)] {Mullaney2013}  Mullaney, J. R., Alexander, D. M., Fine, S., Goulding, A. D., Harrison, C. M., Hickox, R. C. 2013, \mnras, 433, 622

\bibitem[Nelson \& Whittle (1996)] {Nelson1996} Nelson, C. H. \& Whittle, M. 1996, \apj, 465, 96

\bibitem[Nesvadba et al. (2008)] {Nesvadba2008} Nesvadba, N. P. H., Lehnert, M. D., De Breuck, C., Gilbert, A. M., \& van Breugel, W. 2008, A\&A, 491, 407
\bibitem[Nesvadba et al. (2017)] {Nesvadba2017} Nesvadba, N. P. H., De Breuck, C., Lehnert, M. D.,  Best, P. N. \&  Collet, C. 2017, A\&A, 599, 123


\bibitem[Oh et al. (2015)] {Oh2015} Oh, K., Yi, S.K., Schawinski, K., Koss, M., Trakhtenbrot, B., Soto, K. 2015, \apjs, 219, 10
\bibitem[Osterbrock \&  Ferland (2006)] {Osterbrock2006} Osterbrock, D., Ferland, G., 2006, Astrophysics of gaseous nebulae and active galactic nuclei, Sausalito, CA: University Science Book
\bibitem[Popovi\' c et al.(2004)]{popovic2004} Popovi\' c, L.\v C., Mediavilla, E., Bon, E. and Ili\' c, D. 2004, \aap, 423, 909
\bibitem[Rakshit \& Woo (2018)] {Rakshit2018} Rakshit, S. \& Woo, J.-H. 2018, \apj, 865, 5



\bibitem[Sarzi et al. (2006)]{Sarzi2006} Sarzi M. et al., 2006, \mnras, 366, 1151

\bibitem[Schlafly \& Finkbeiner (2011)]{SF2011} Schlafly, E. F., Finkbeiner, D. P. 2011, \apj, 737, 103
\bibitem[Sexton et al. (2021)] {Sexton2021} Sexton, R. O., Matzko, W., Darden, N., Canalizo, G., Gorjian, V. 2021, \mnras, 500, 2871

\bibitem[Smee et al. (2013)]{SM2013} Smee, S. A., Gunn, J. E., Uomoto, A., Roe, N. et al. \ 2013, \aj, 146, 32
\bibitem[Smirnova et al. (2007)]{Smirnova2007} Smirnova, A.A., Gavrilovi\'{c}, N., Moiseev, A.V., Popovi\'{c}, L. \v{C}. et al. \ 2007, \mnras, 377, 480
\bibitem[Thomas et al. (2013)]{Thomas2013} Thomas, D.,  Steele, O., Maraston, C., Johansson, J. et al.  2013, \mnras, 431, 1383


\bibitem[Tombesi et al. (2015)]{Tombesi2015} Tombesi F., Mel\'{e}ndez M., Veilleux S., Reeves J. N., Gonz\'{a}lez-Alfonso E.,
Reynolds C. S. 2015, Nature, 519, 436

\bibitem[Trippe et al. (2010)]{Trippe2010} Trippe, M. L., Crenshaw, D. M., Deo, R. P., Dietrich, M., Kraemer, S. B., Rafter, S. E., Turner, T. J. \ 2010, \apj, 725, 1749


\bibitem[Vanden Berk et al.(2006)]{2006AJ...131..84} Vanden Berk, D.~E., Shen, J., Yip, C.-W. et al.\ 2006, \aj, 131, 84

\bibitem[Veilleux et al. (2005)]{Veilleux2005} Veilleux, S., Cecil, G., Bland-Hawthorn, J. 2005,  ARA\&A, 43, 769

\bibitem[Venturi et al. (2021)]{Venturi2021} Venturi, G., Cresci, G., Marconi. A., Mingozzi, M. et al. 2021,  A\&A, 648, 17

\bibitem[Wang et al. (2018)]{Wang2018} Wang, J., Xu, D. W. \&  Wei, J. Y. 2018, \apj, 852, 26
\bibitem[Westmoquette et al.(2007)]{Westmoquette2007} Westmoquette,  M. S., Smith, L. J.,  Gallagher, J. S., Exter, K. M.  2007, \mnras, 381, 913

\bibitem[Woo et al.(2014)]{woo2014} Woo, J.-H., Kim, J.-G., Park, D., Bae, H.-J., et al.  2014, JKAS, 47, 167

\bibitem[Woo et al.(2016)]{woo2016} Woo, J.-H., Bae, H.-J., Son, D., Karouzos, M. 2016, \apj, 817, 108
\bibitem[Woo et al.(2017)]{woo2017} Woo, J.-H., Son, D. \& Bae, H.-J.  2017, \apj, 839, 120

\bibitem[Xiao et al.(2011)]{Xiao2011} Xiao, T., Barth, A. J., Greene, J. E., Ho, L. C.  et al. 2011,  \apj, 739, 28

\bibitem[Yip et al.(2004a)]{yip04a} Yip, C.~W., Connolly, A.~J., Szalay, A.~S. et al.\ 2004a, \aj, 128, 585
\bibitem[Yip et al.(2004b)]{yip04b} Yip, C.~W., Connolly, A.~J., Vanden Berk, D.~E. et al. 2004b, \aj, 128, 2603
\bibitem[Zakamska \& Greene (2014)]{Zakamska2014} Zakamska, N. L. \& Greene, J. E. 2014, \mnras, 442, 784
\bibitem[Zakamska et al. (2016)]{Zakamska2016} Zakamska, N. L., Hamann, F., P\^ aris, I, Brandt, W. N. et al. 2016, \mnras, 459, 3144

\bibitem[Zhang et al. (2011)]{Zhang2011} Zhang, K.,Dong, X.-B., Wang, T.-G. \& Gaskell, M. C. 2011, \apj, 737, 71

\end{thebibliography}
\end{document}